\documentclass[10pt,letterpaper]{article}
	\usepackage[top=0.85in,left=2.75in,footskip=0.75in]{geometry}
	\usepackage{amsmath,amssymb}
	\usepackage[utf8x]{inputenc}
	\usepackage{caption}
	\usepackage{cite}
	\raggedright
	\usepackage[aboveskip=1pt,labelfont=bf,labelsep=period,justification=raggedright,singlelinecheck=off]{caption}
	
	\makeatletter
	\renewcommand{\@biblabel}[1]{\quad#1.}
	\makeatother
	\usepackage{lastpage,fancyhdr,graphicx}
	\usepackage{epstopdf}
	\fancyheadoffset[L]{2.25in}
	\fancyfootoffset[L]{2.25in}
	\newcommand\twofigMAT{0.49\textwidth}
	
	\newcommand\onefigMAT{0.7\textwidth}
	
	\usepackage{float}

\begin{document}
	\vspace*{0.2in}
	\begin{flushleft}
    {\Large
     \textbf\newline{\textbf{Threshold-activated transport stabilizes chaotic 
     populations to steady states}} 
% Please use "sentence case" for title and headings (capitalize only the first 
%word in a title (or heading), the first word in a subtitle (or subheading), 
%and 
%any proper nouns).
     }	
	\newline
	% Insert author names, affiliations and corresponding author email (do not 
	%include titles, positions, or degrees).
	\\
	\textbf{Chandrakala Meena\textsuperscript{},
 	Pranay Deep Rungta\textsuperscript{},
	Sudeshna Sinha\textsuperscript{*}}

	\bigskip
	\textbf{}Department of Physical Sciences, Indian Institute of Science  
	Education and Research (IISER) Mohali, Knowledge City, SAS Nagar,
  Sector 81, Manauli PO 140306, Punjab, India. \\
  
	*sudeshna@iisermohali.ac.in
	
	\end{flushleft}
	\section*{Abstract}

	We explore Random Scale-Free networks of populations, modelled by chaotic Ricker maps, connected by transport
	that is triggered when population density in a patch is in excess of a critical threshold level. Our central result
	is that threshold-activated dispersal leads to stable fixed populations, for a wide range of threshold levels.
	Further, suppression of chaos is facilitated when the threshold-activated migration is more rapid than the intrinsic
	population dynamics of a patch. Additionally, networks with large number of nodes open to the environment, readily yield
	stable steady states. Lastly we demonstrate that in networks with very few open nodes, the degree and betweeness centrality
	of the node open to the environment has a pronounced influence on control. All qualitative trends are corroborated by
	quantitative measures, reflecting the efficiency of control, and the width of the steady state window.

	\section*{Introduction}

	Nonlinear systems, describing both natural phenomena as well as human-engineered devices, can give rise to a rich gamut
	of patterns ranging from fixed points to cycles and chaos. So mechanisms that enable a chaotic system to maintain a fixed
	desired activity (the ``goal'') has witnessed enormous research attention \cite{control}. In early years the focus was on
	controlling low-dimensional chaotic systems, and guiding chaotic states to desired target states \cite{adaptive,ogy,glass}.
	Efforts then moved on to the arena of lattices modelling extended systems, and the control of spatiotemporal patterns in
	such systems \cite{pattern_control}. With the	advent of network science to describe connections between complex sub-systems,
	the new challenge is to find mechanisms or strategies that are capable of stabilizing these large interactive systems \cite{motter}. 

	In this work we consider a network of population patches \cite{metapopulation_review,meta2}, or ``a population of populations'' \cite{levin2}.
	Now, most models of metapopulation dynamics consider density dependent dispersal, analogous to reaction-diffusion processes
	\cite{colizza,dispersal2,dispersal4,dispersal5,dispersal6}. Here we consider a different scenario, namely one where the inter-patch
	connection is triggered by the excess of population density in a patch \cite{glass,ssprl}. This describes a system comprising of many
	 spatially discrete sub-populations connected by threshold-activated dispersal. Our principal question will be the following:
	  {\em can threshold-activated coupling serve to control networks of intrinsically chaotic populations on to regular behaviour? }

	In the sections below, we will first discuss details of the nodal dynamics, as well as the salient features of pulsatile transport
	triggered by threshold mechanisms. We will then go on to demonstrate, through qualitative and quantitative measures, that such
	threshold-activated connections manage to stabilize chaotic populations to steady states. Further we will explore how the
	 critical threshold that triggers the migration, and the timescales of the nodal dynamics vis-a-vis transport, influences
	 the emergent dynamics.\\
	
	\section*{Model}

	Consider a network of $N$ sub-systems, characterized by variable $x_n(i)$ at each node/site $i$ ($i=1,\dots N$) at time instant $n$.
	 Specifically, we study a prototypical map, the Ricker (Exponential) Map, at the local nodes. Such a map models population growth of
	  species with non-overlapping generations, and is given by the functional form:	

	\begin{equation}
		x_{n+1} (i) =f(x_{n} (i))= x_n(i) \exp(r(1-x_n(i)))
	\end{equation}
	
	where $r$ is interpreted as an intrinsic growth rate and (dimensionless) $x_n(i)$ is the population scaled by the carrying capacity at
	generation $n$ at node/site $i$. We consider $r=4$ in this work, namely, an isolated uncoupled population patch displays {\em chaotic} behaviour.
	
	The coupling in the system is triggered by a threshold mechanisms \cite{glass,sspre,kazu}. Namely, the dynamics of node $i$ is such that
	 if $x_{n+1} (i) > x_c$, the variable is adjusted back to $x_c$ and the ``excess'' $x_{n+1}-x_c$ is distributed to the neighbouring
	  patches. The threshold parameter $x_c$ is the critical value the state variable has to exceed in order to initiate threshold-activated
	   coupling. So this class of coupling is {\em pulsatile}, rather than the more usual continuous coupling forms, as it is triggered
	    {\em only} when a node exceeds threshold.
	
		Specifically, we study such population patches coupled in a Random Scale-Free network, where the network of underlying connections
		is constructed via the Barabasi-Albert preferential attachment algorithm, with the number of links of each new node denoted by parameter
		$m$ \cite{scalefree}. The resultant network is characterized by a fat-tailed degree distribution, found widely in nature.
		The underlying web of connections determines the ``neighbours'' to which the excess is equi-distributed. Further, certain nodes
		in the network may be open to the environment, and the excess from such nodes is transported out of the system.
	    Such a scenario will model an open system, and such nodes are analogous to the ``open edge of the system''. 
		We denote the fraction of open nodes in the network, that is the number of open nodes scaled by system size $N$,
		by $f^{open}$. In this work we also consider closed systems with no nodes open to the environment, where nothing is
		transported out of the system, i.e. $f^{open}=0$.
	
		The threshold-activated migration from an over-critical patch can trigger subsequent transport, as the redistribution of excess
		can cause neighbouring sites to become over-critical, thus initiating a domino effect, much like an ``avalanche'' in models of
		self-organized criticality \cite{bak}. All transport within patches stop when all patches are under the critical value, i.e. all
		 $x (i) < x_c$. So there are two natural time-scales here. One time-scale characterizes the chaotic update of the populations at
		 node $i$. The other time scale involves the redistribution of population densities arising from threshold-activated transport.
		 We denote the time interval between chaotic updates, namely the time available for redistribution of excess resulting from 
		 threshold-activated transport processes, by $T_R$. This is analogous to the {\em relaxation time} in models of self-organized 
		 criticality, such as the influential sandpile model \cite{bak}. $T_R$ then indicates the comparative time-scales of the 
		 threshold-activated migration and the intrinsic population dynamics of a patch.\\

	%\bigskip
	\section*{Results}

	We have simulated this threshold-coupled scale-free network of populations, under varying threshold levels
	$x_c$ ($0 \le x_c \le 2$). We considered networks with varying number of open nodes, namely systems that have
	different nodes/sites open to the environment from where the excess population can migrate out of the system.
	Further, we have studied a range of redistribution times $T_R$, capturing different timescales of migration vis-a-vis
	population change \cite{relax}. With no loss of generality, in the sections below, we will present salient results for Random Scale-Free
	networks with $m=1$, and specifically demonstrate, both qualitatively and quantitatively, the stabilization of networks of 
	chaotic populations to steady-states under threshold-activated coupling.

	\subsection*{Emergence of Steady States}
	
     First, we consider the case of large $T_R$, where the transport processes 
     are fast compared to the population dynamics, or equivalently, the 
     population dynamics of the patch is slow compared to inter-patch 
     migrations. Namely, since the chaotic update is much slower than the 
     transport between nodes, the situation is analogous to the slow driving 
     limit \cite{bak}. In such a case, the system has time for many transport 
     events to occur between chaotic updates, and avalanches can die down, i.e. 
     the system is ``relaxed'' or ``under-critical'' between the chaotic 
     updates. So when the transport/migration is significantly faster than the 
     population update (namely the time between generations), the system tends 
     to reach a stationary state where all nodal populations are less than 
     critical. 

\begin{figure}[H]
	\centering 
		\includegraphics[trim={2.5cm 0 2cm 0},clip,width=\twofigMAT]{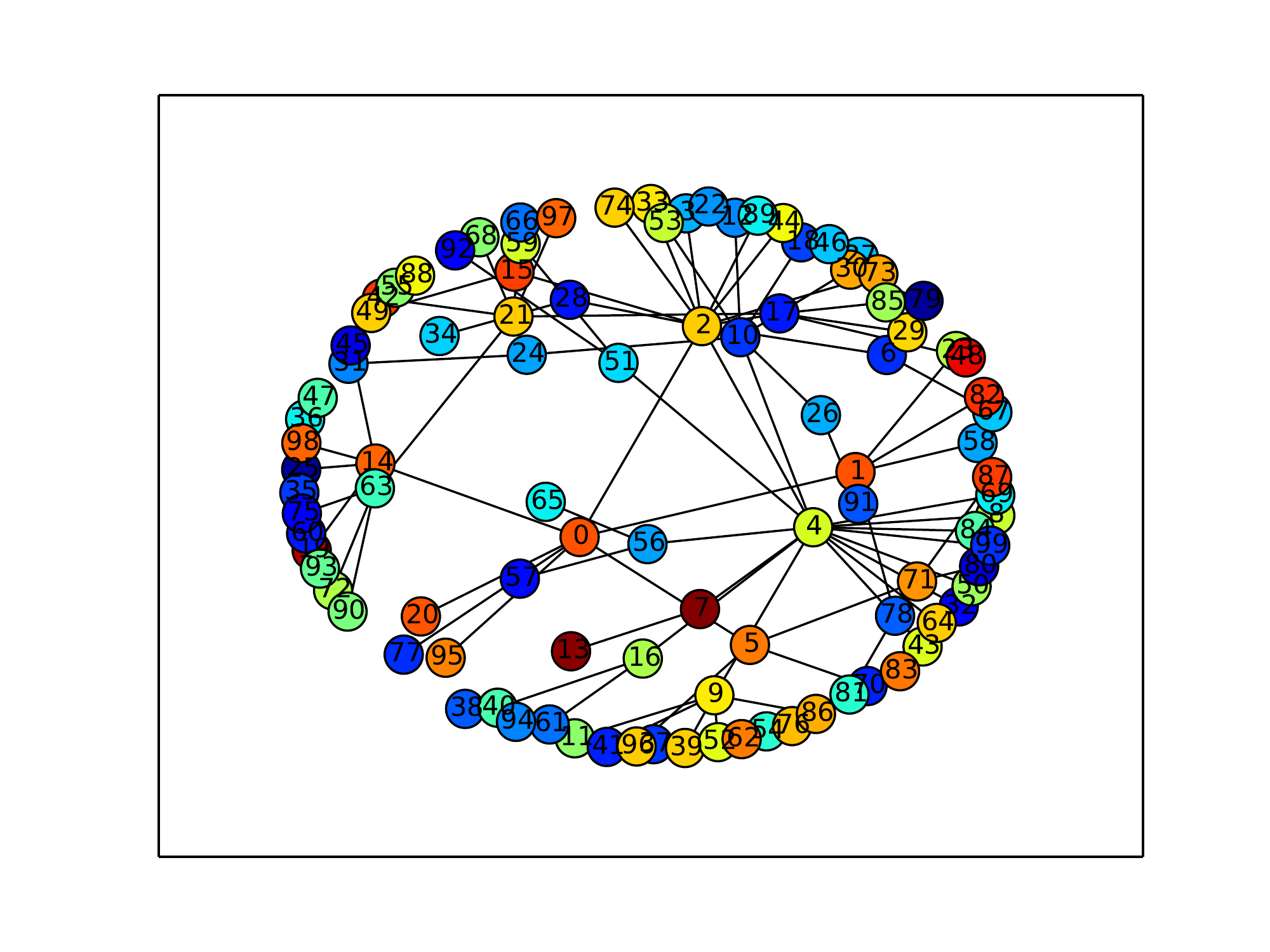}
		\includegraphics[trim={2.5cm 0 2cm 0},clip,width=\twofigMAT]{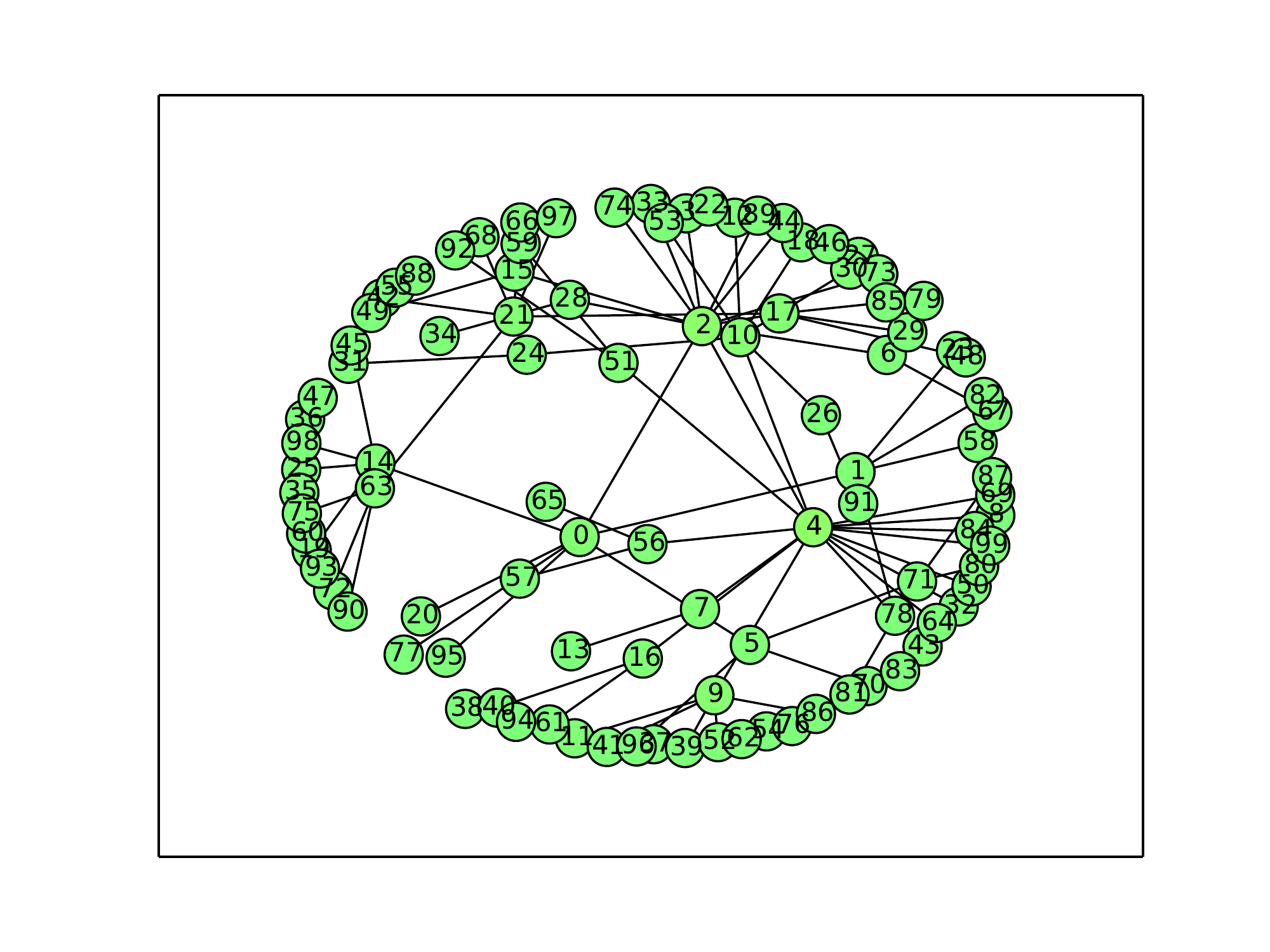}
		\caption{ \textbf{State of the nodes (coded in color) in a Random Scale-Free 
		Network of intrinsically chaotic populations under threshold-activated 
		coupling, at different instants of time.} Here the steady state 
		value represented by the light green color. The left panel displays the 
		network at initial time, showing the random initial state of the 
		network. The right panel shows the network after $50$ time steps, 
		clearly showing that all nodes have evolved to a steady state (as 
		evident from the uniform light green color). Here redistribution time 
		$T_R=5000$ and the critical threshold $x_c=0.5$, and there is a single 
		node open to the environment.} \label{nodes}
	\end{figure}

	An illustrative case of the state of the nodes in the network is shown in 
	Fig.~\ref{nodes}. Without much loss of generality, we display results for a 
	network of size $N=100$, for a representative large value of redistribution 
	time $T_R=5000$. It is clear that {\em all the nodes in the network gets 
	stabilized to a fixed point}, namely all population patches evolve to a 
	stable steady state. 

\begin{figure}[H]
	\centering 
		\includegraphics[trim={0.2cm 0.1cm 0.2cm 0.1cm},clip,width=\twofigMAT]{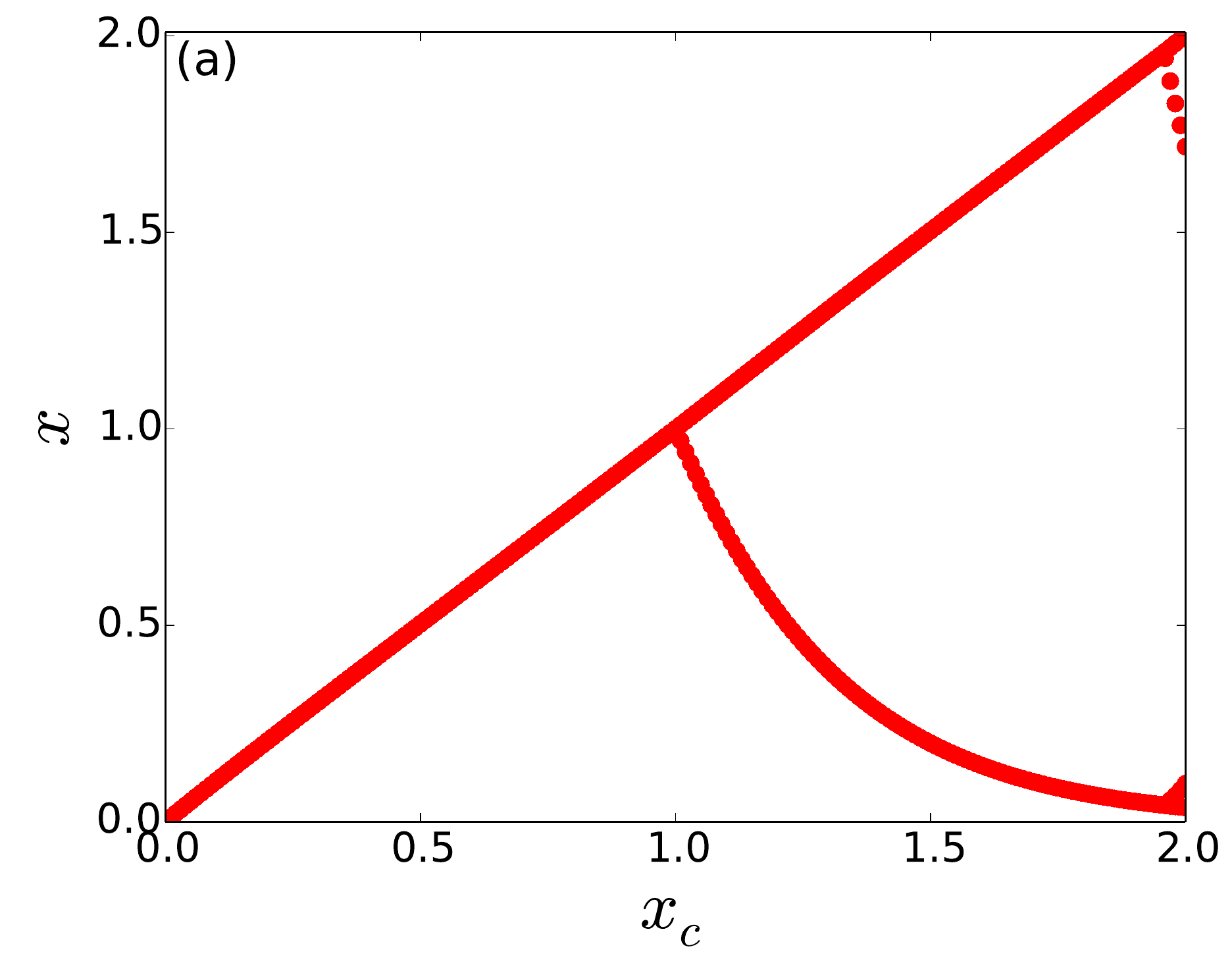}
		\includegraphics[trim={0.2cm 0.1cm 0.2cm 0.1cm},clip,width=\twofigMAT]{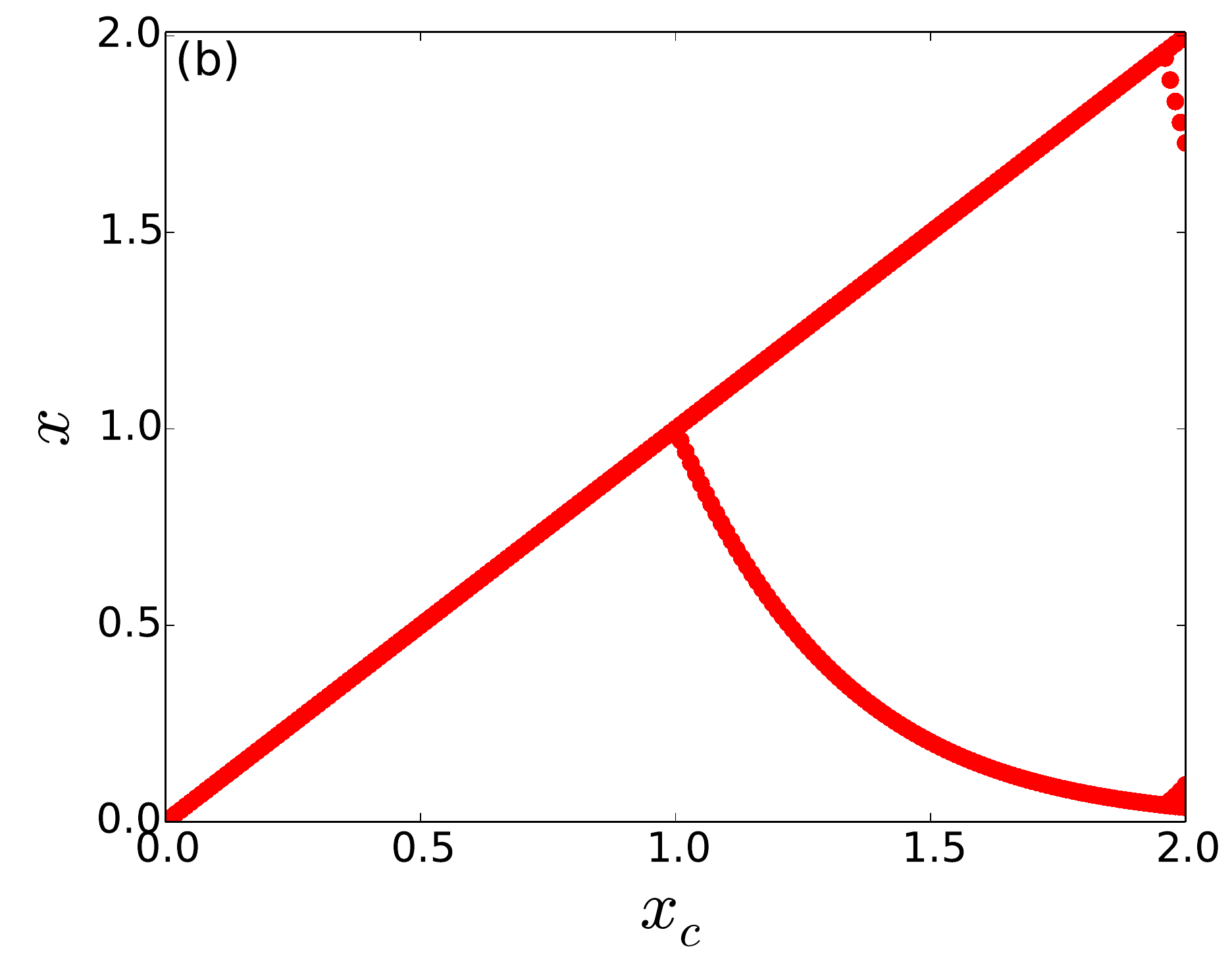}
		\caption{\textbf{Bifurcation diagrams of the state of a representative node, 
		with respect to critical threshold $x_c$, in a threshold-coupled Random 
		Scale-Free network of intrinsically chaotic populations.} Here 
		$T_R=5000$ and the network has a single open node, of degree (a) $1$ and (b) $15$.}\label{withedgeRandomSFm1highT_R}
	\end{figure}

	The next natural question is the influence of the critical threshold $x_c$ on 
	the emergent dynamics. This dependence is demonstrated in bifurcation diagrams 
	displayed in Fig.~\ref{withedgeRandomSFm1highT_R}. It is clearly evident from 
	these that a {\em large window of threshold values} ($0 \le x_c < 1$) yield 
	spatiotemporal steady states in the network \cite{thresh}. It is also apparent 
	that the degree of the open node does not affect the emergence of steady states 
	here. Further, for threshold values beyond the window of control to fixed 
	states, one obtains cycles of period $2$. Namely for threshold levels $1 < x_c 
	< 2$ the populations evolve in regular cycles, where low population densities 
	alternate with a high population densities. This behaviour is reminiscent of 
	the field experiment conducted by Scheffer et al \cite{self-perpetuating} which 
	showed the existence of self-perpetuating stable states alternating between 
	blue-green algae and green algae. We discuss the underlying reason for this 
	behaviour in the Appendix, and offer analytical reasons for the range of 
	period-$1$ and $2$ behaviour considering a single threshold-limited map.
		
	So our first result can be summarized as follows: when redistribution time 
	$T_R$ is large and the critical threshold $x_c$ is small, we have very 
	efficient control of networks of chaotic populations to steady states. This 
	suppression of chaos and quick evolution to a stable steady states occurs 
	irrespective of the number of open nodes.

	\setlength{\belowcaptionskip}{2cm}
	\setlength{\belowcaptionskip}{0.2cm}	
  \subsection*{Influence of the redistribution time and the number of open 
  nodes on the suppression of chaos}

	Now we focus on the network dynamics when $T_R$ is small, and the 
	time-scales of the nodal population dynamics and the inter-patch transport 
	are comparable. So now there will be nodes that may remain over-critical at 
	the time of the subsequent chaotic update, as the system does not have 
	sufficient time to ``relax'' between population updates. The network is 
	then akin to a rapidly driven system, with the de-stabilizing effect of the 
	chaotic population dynamics competing with the stabilizing influence of the 
	threshold-activated coupling. So for small $T_R$, the system does not get 
	enough time to relax to under-critical states and so perfect control to 
	steady states may not be achieved.

\begin{figure}[!h]
         \centering                        
         \includegraphics[trim={1.5cm 0.5cm 3cm 1cm},clip,width=\twofigMAT]{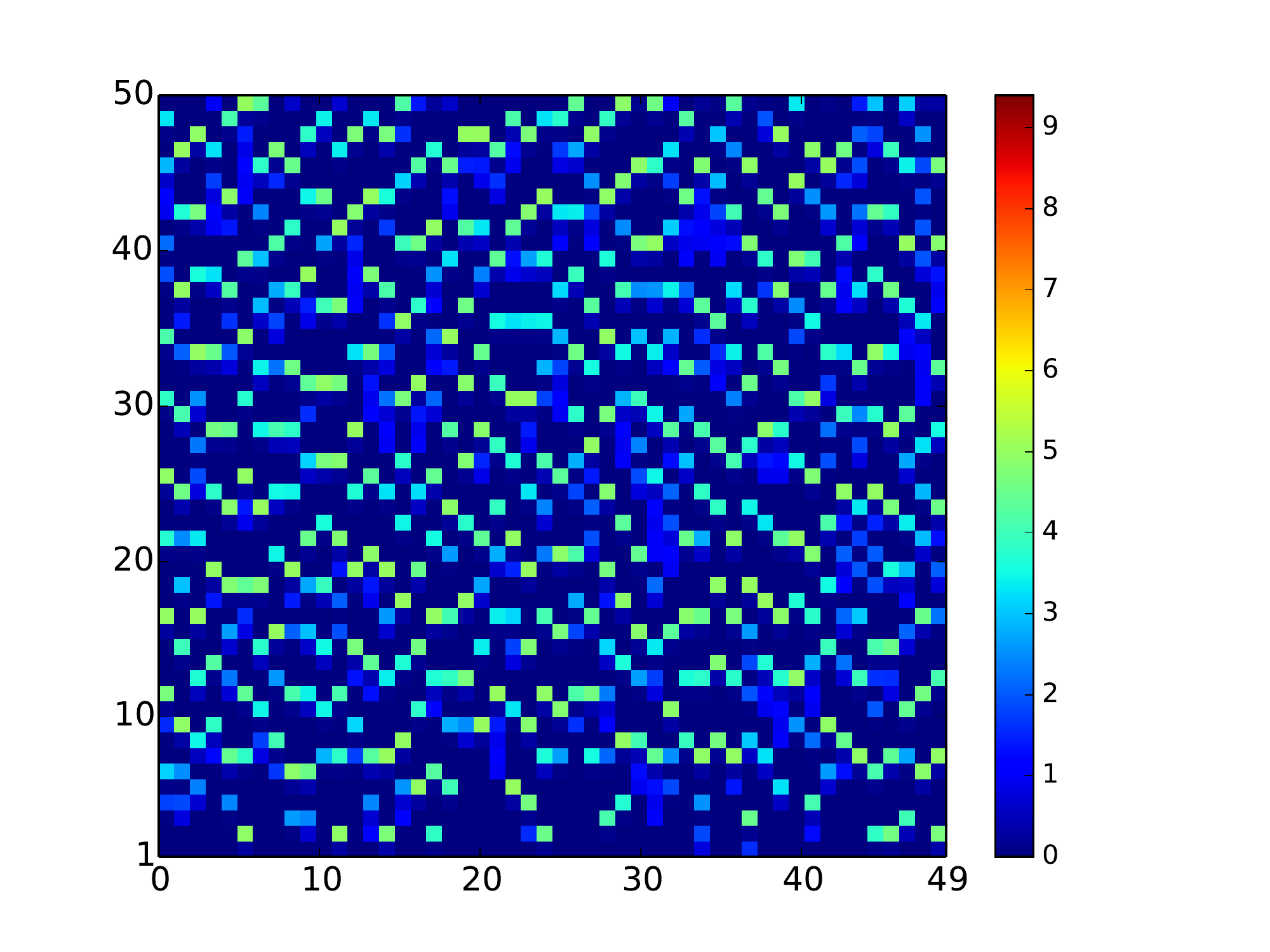}
         \includegraphics[trim={1.5cm 0.5cm 3cm 1cm},clip,width=\twofigMAT]{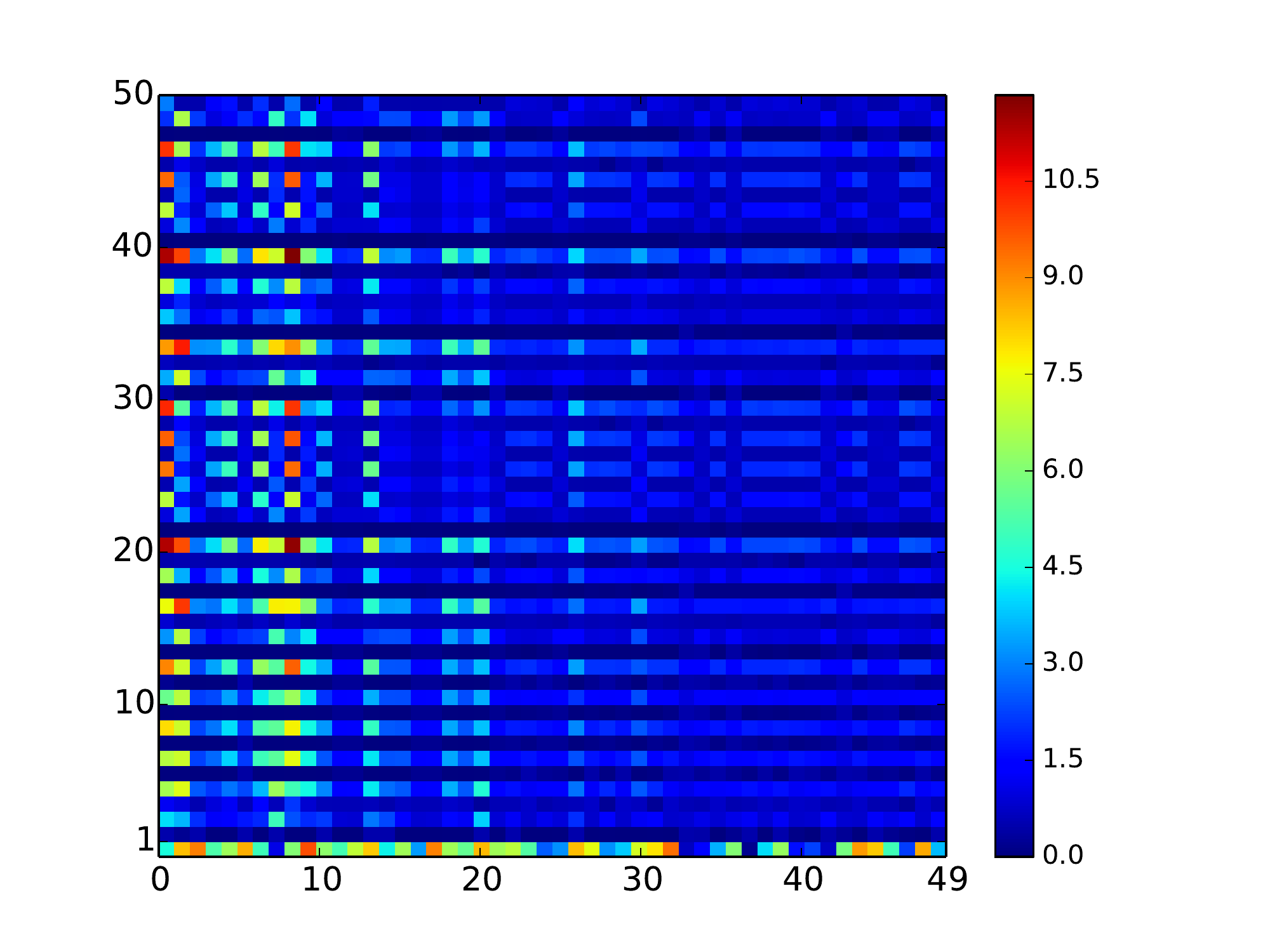}
         \includegraphics[trim={1.5cm 0.5cm 3cm 1cm},clip,width=\twofigMAT]{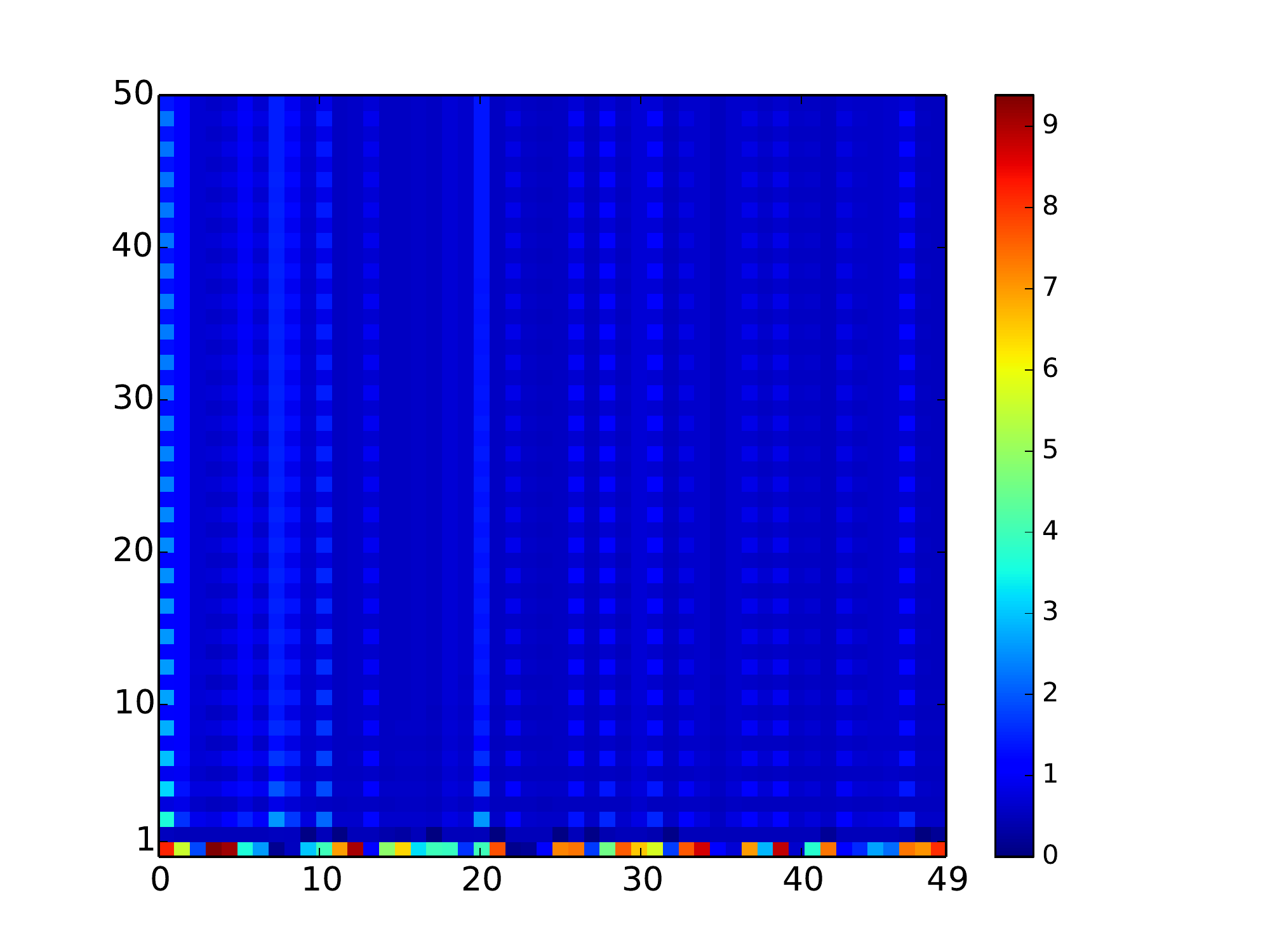}
         \includegraphics[trim={1.5cm 0.5cm 3cm 1cm},clip,width=\twofigMAT]{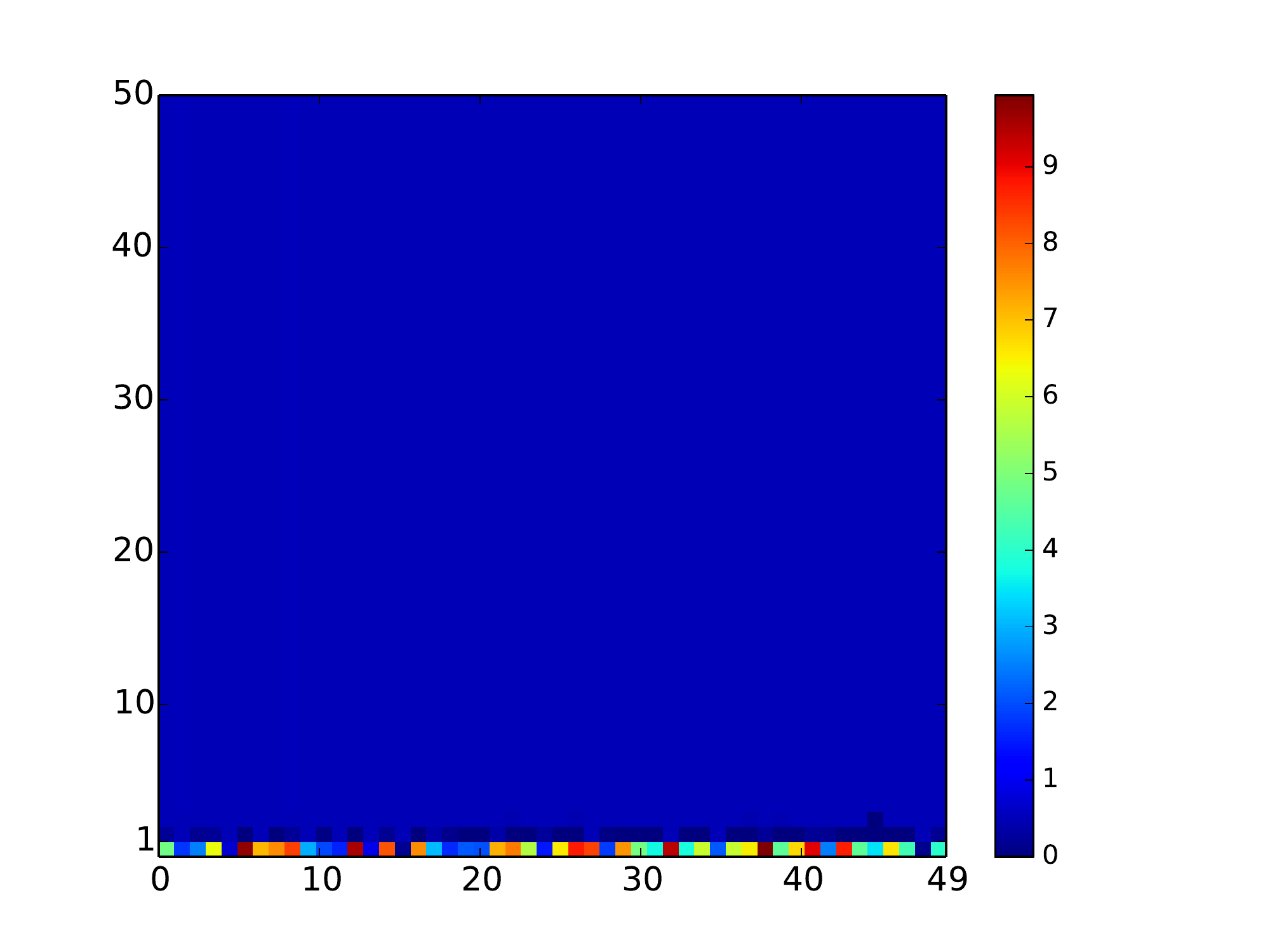}
         \caption{\textbf{Space-time plots displaying the spatiotemporal 
         behaviour of a Random Scale-Free network of intrinsically chaotic populations.}
         Here time runs along the vertical axis and site index displayed along the 
         horizontal axis. Panel (a) shows the 
         case of uncoupled chaotic populations evolving from a 
         representative random initial state. Panels (b), (c) and 
         (d) show the evolution of the same populations connected 
         through threshold-activated coupling. System size $N=50$, 
         redistribution time $T_R=50$, the critical threshold 
         $x_c=0.5$ and the number of open nodes in the network is 
         (b) $1$, (c) $10$, (d) $30$.} \label{spacetimeRSFm1}
     \end{figure}

\begin{figure}[!h]
	\centering
	\includegraphics[width=\twofigMAT]{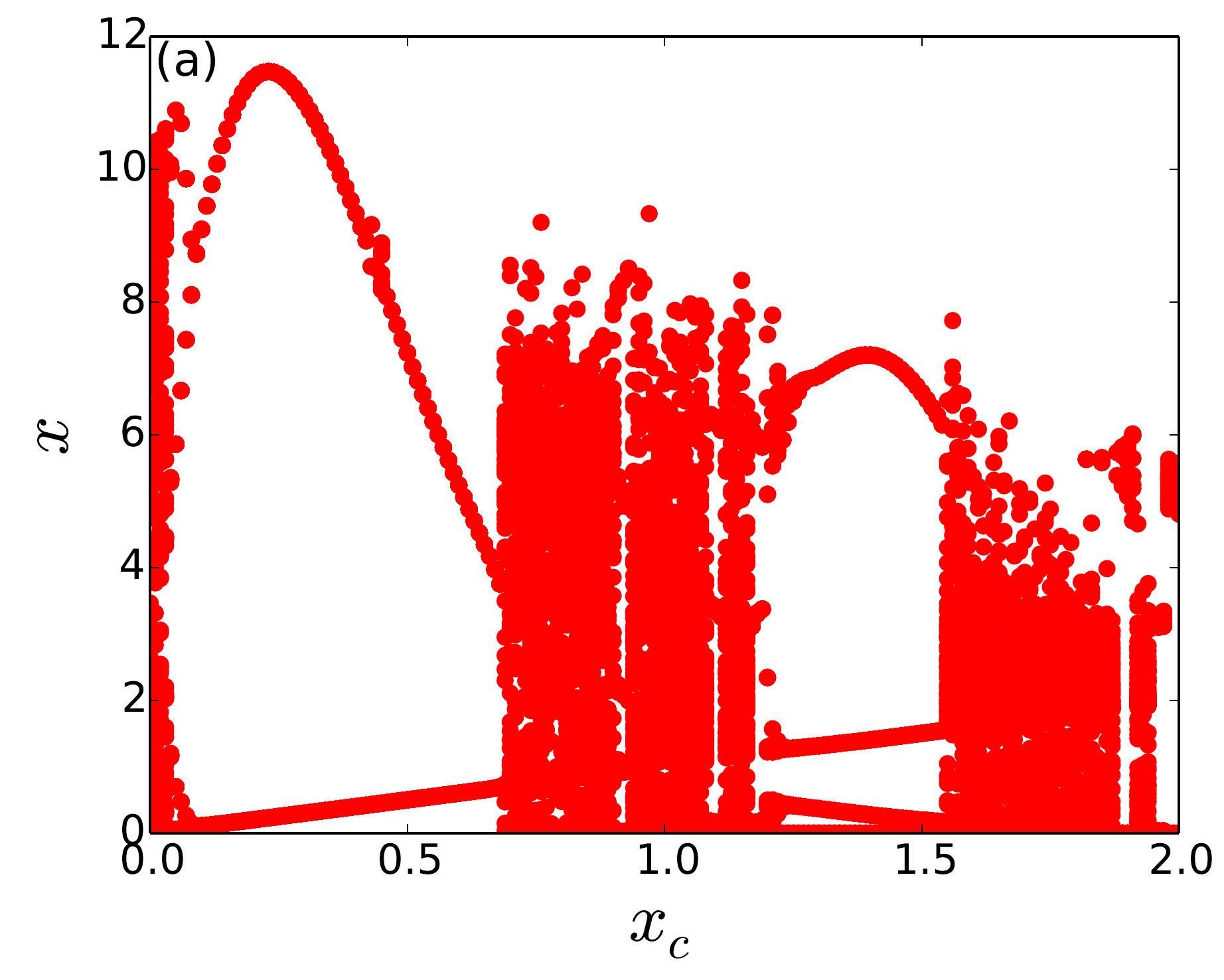}
	\includegraphics[width=\twofigMAT]{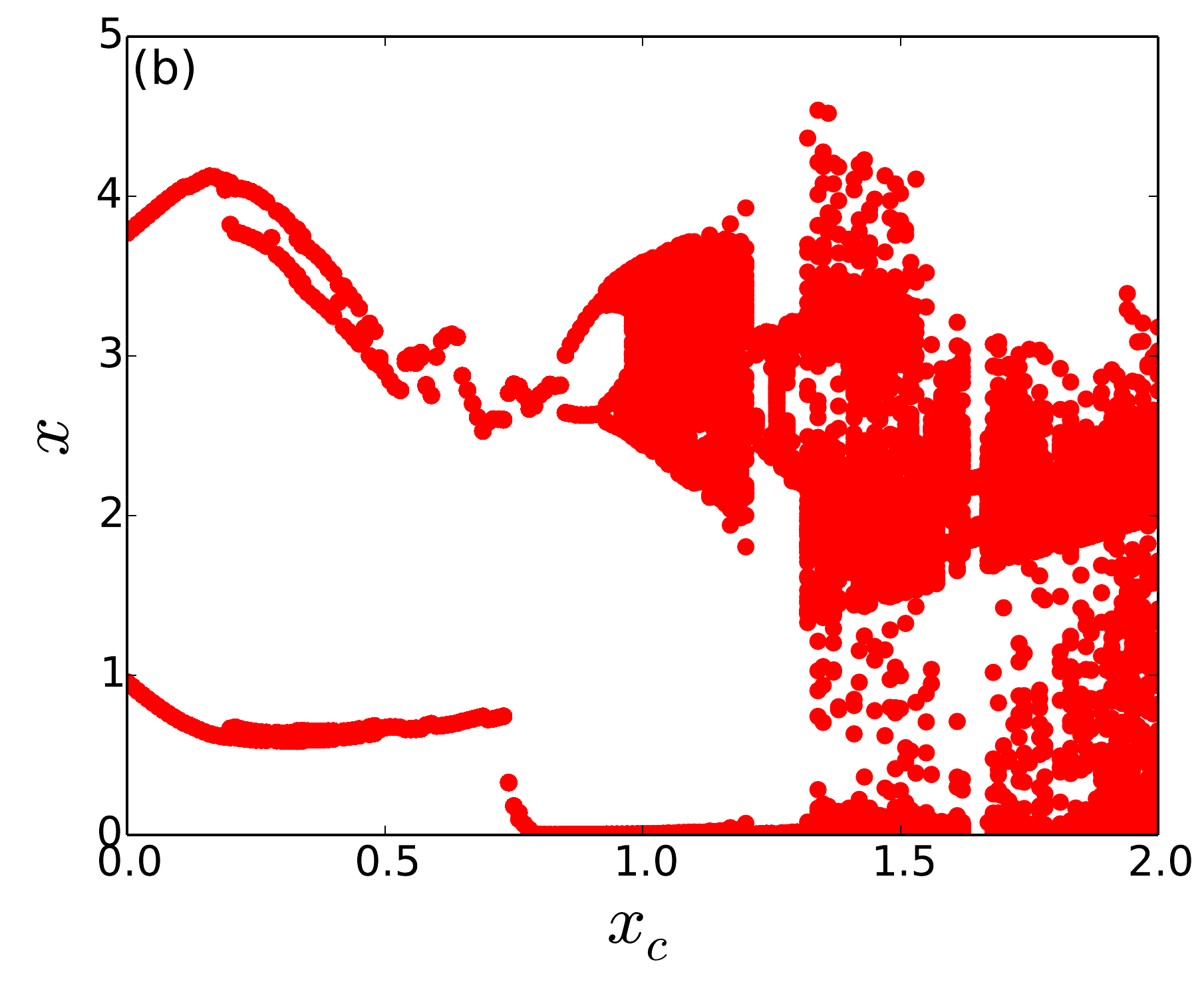}
 	\includegraphics[width=\twofigMAT]{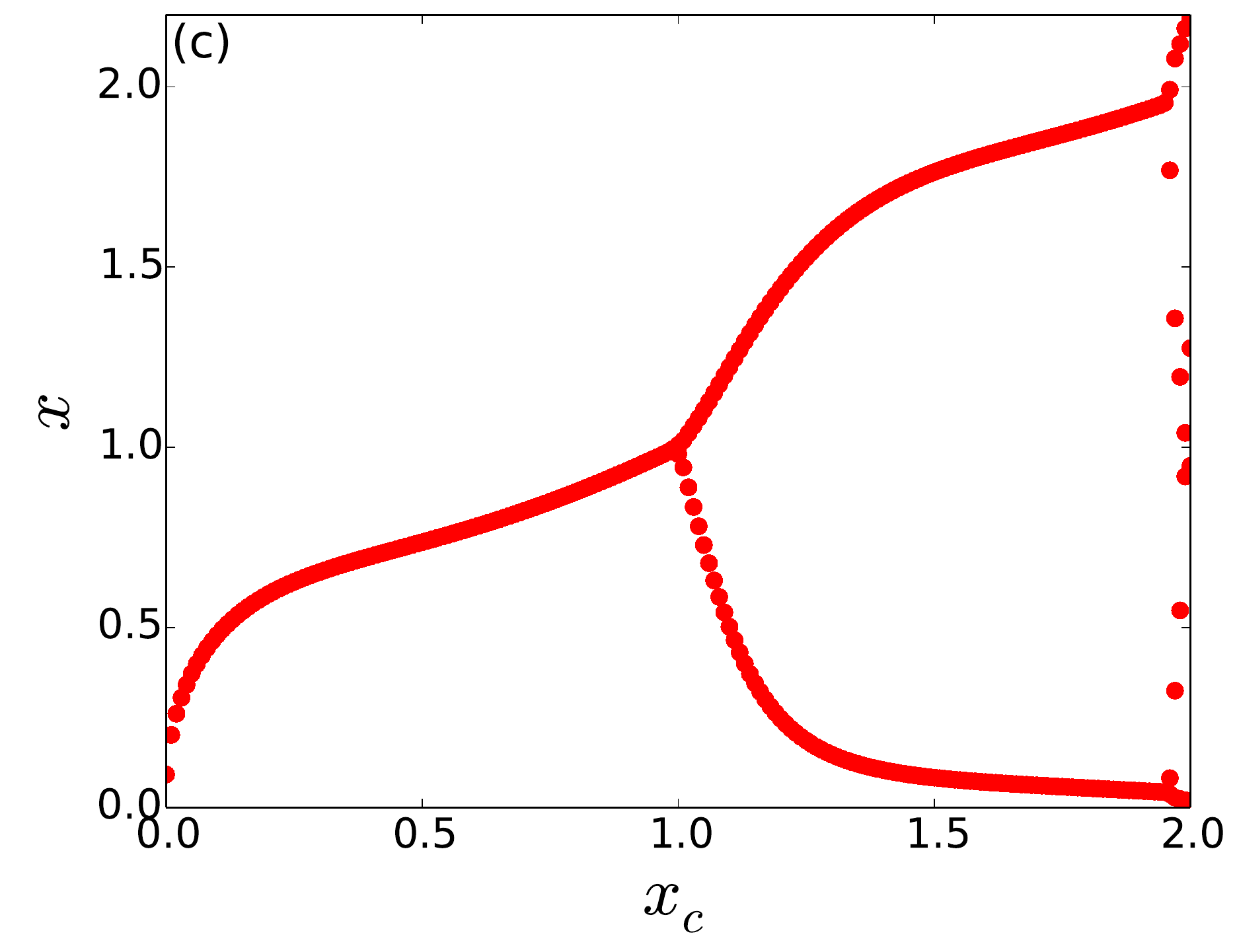}
 	\includegraphics[width=\twofigMAT]{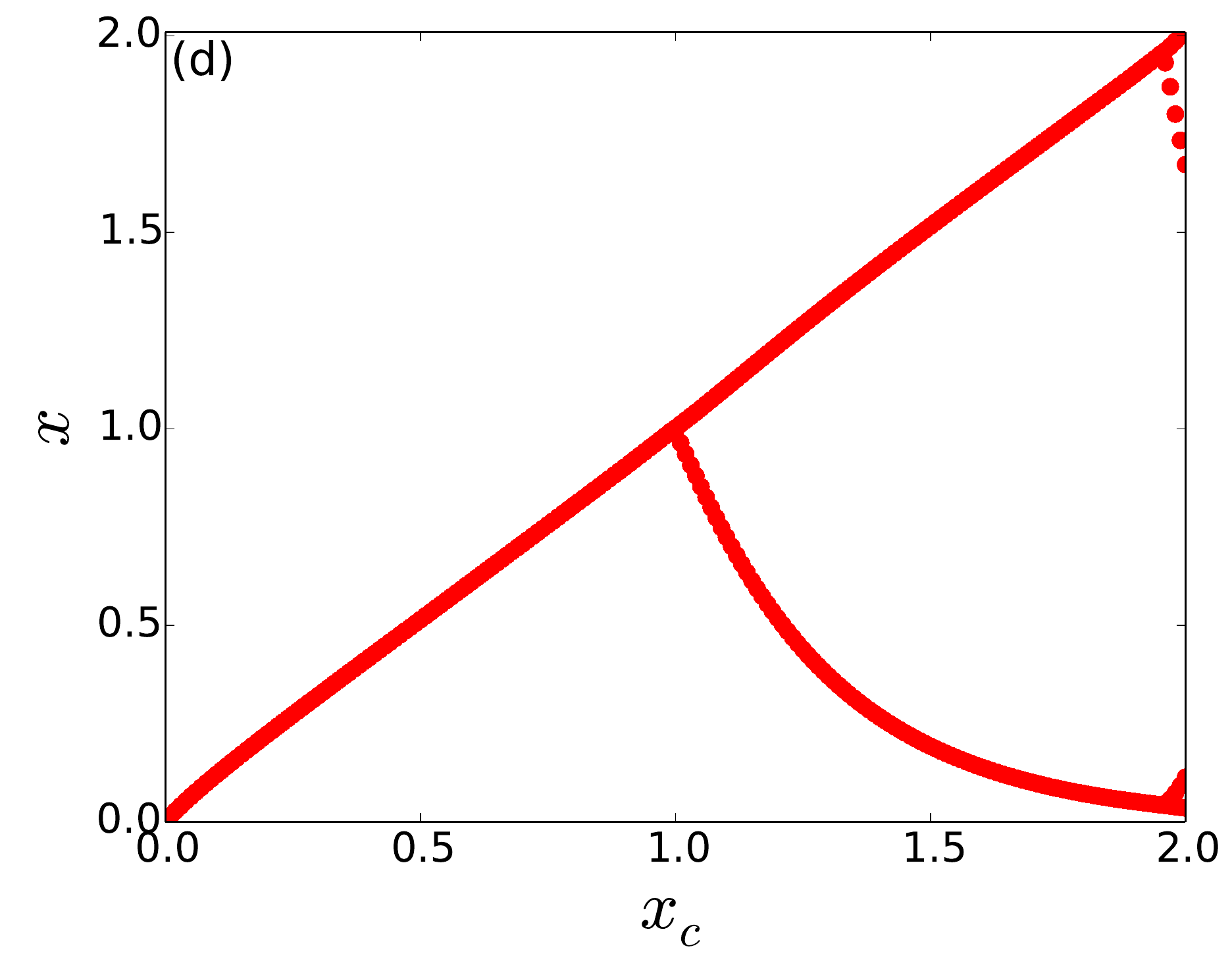}
	\caption{\textbf{Bifurcation diagrams for one representative node in a 
	threshold-coupled Random Scale-Free network of intrinsically chaotic 
	populations, with respect to critical threshold $x_c$.} Here $T_R=50$ 
	and the number of open nodes is (a) $1$, (b) $10$, (c) $30$ and (d) $60$.}
	\label{withedgeRandomSFm1lowT_R}
	\end{figure}

     Importantly now, the fraction of open nodes $f^{open}$ is crucial to chaos 
     suppression. In general, a larger fraction of open nodes facilitates 
     control of the intrinsic chaos of the nodal population dynamics, as the 
     de-stabilizing ``excess'' is transported out of the system more 
     efficiently. We investigate this dependence, through space-time plots of 
     representative networks with varying number of open nodes and 
     redistribution times (cf. Fig.~\ref{spacetimeRSFm1}), and through 
     bifurcation diagrams of this system with respect to critical threshold 
     $x_c$ (cf. Fig.~\ref{withedgeRandomSFm1lowT_R}).

	It is apparent from Fig.~\ref{spacetimeRSFm1}, that when there are enough 
	open nodes, the network relaxes to the steady state even for low 
	redistribution times. Also notice from Fig.~\ref{spacetimeRSFm1}(d) that 
	the system {\em reaches the steady state very rapidly}, namely within a few 
	time steps, from the random initial state. So more open nodes yields better 
	control of the intrinsic chaos of the nodal population dynamics to fixed 
	populations. This is also corroborated in the bifurcation diagrams 
	displayed in Fig.~\ref{withedgeRandomSFm1lowT_R}, where control to steady 
	states is seen even for low $T_R$, when there are large number of open 
	nodes, vis-a-vis networks with few open nodes. Further contrast this with 
	the dynamics of a system with large $T_R$, shown earlier in  
	Fig.~\ref{withedgeRandomSFm1highT_R}, where even a {\em single} open node 
	leads to stable steady states for a large range of threshold values. 
	Similar qualitative trends are also borne out in Random Scale-Free network 
	with $m=2$, where again more open nodes and longer redistribution times 
	result in better control to fixed population densities.

\begin{figure}[!h]
	\centering 
	\includegraphics[width=\onefigMAT]{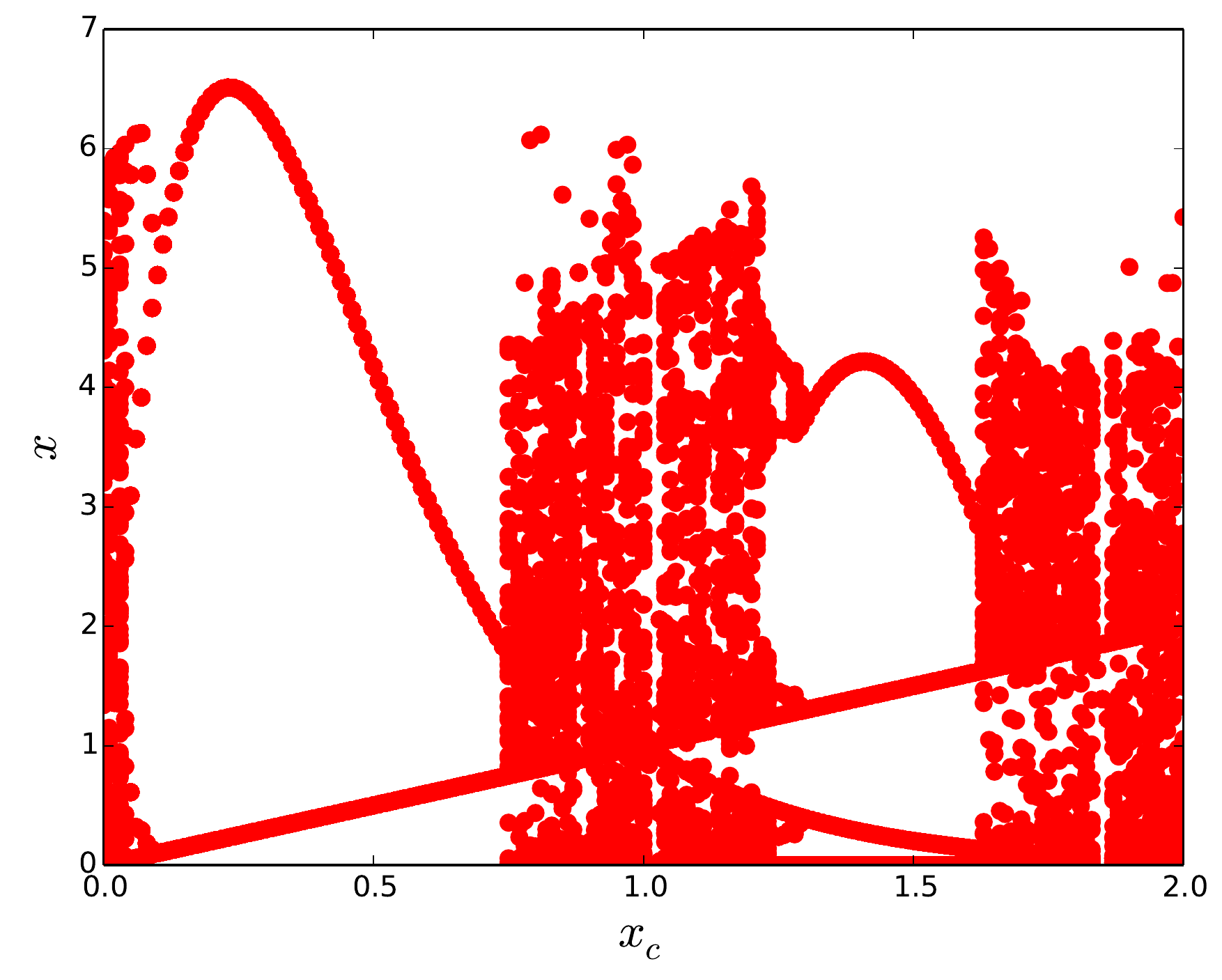}
	\caption{\textbf{Bifurcation diagram displaying the state of a representative 
		site, for threshold-coupled populations in a Random Scale-Free 
		network.} Here $T_R=5000$ and there are no open 
		nodes.}\label{withoutedgeRandomSF1m1}
	\end{figure}

	As a limiting case, we also studied the spatiotemporal behaviour of 
	threshold-coupled networks without open nodes. Here the network of coupled 
	population patches is a closed system. Again the intrinsic chaos of the 
	populations is suppressed to regular behaviour, for large ranges of threshold 
	values. However, rather than steady states, one now obtains period-2 cycles. 
	This is evident through the bifurcation diagram of a closed network (cf. 
	Fig.~\ref{withoutedgeRandomSF1m1}) vis-a-vis networks with at least one open 
	node (cf. Fig.~\ref{withedgeRandomSFm1highT_R}). Also, note the similarity of 
	the bifurcation diagram of the closed system with that of a system with low 
	$T_R$ and few open nodes. This similarity stems from the underlying fact that 
	in both cases the network cannot relax to completely under-critical states by 
	redistribution of excess between the population updates, either due to paucity 
	of time for redistribution (namely low $T_R$) or due to the absence of open 
	nodes to transport excess out of the system.
			
\begin{figure}[!h]
     \centering
   	 \includegraphics[width=\twofigMAT]{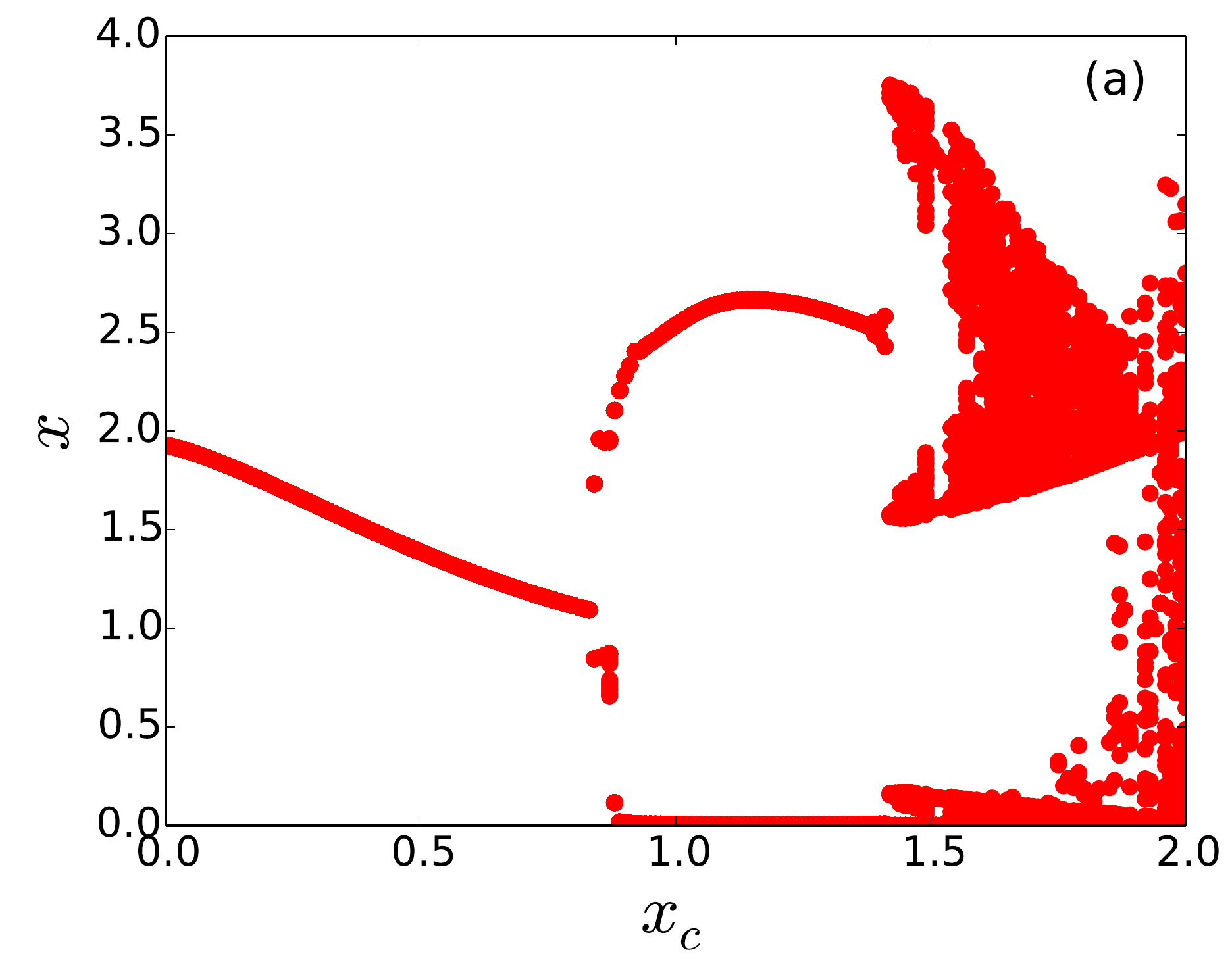}
    \includegraphics[width=\twofigMAT]{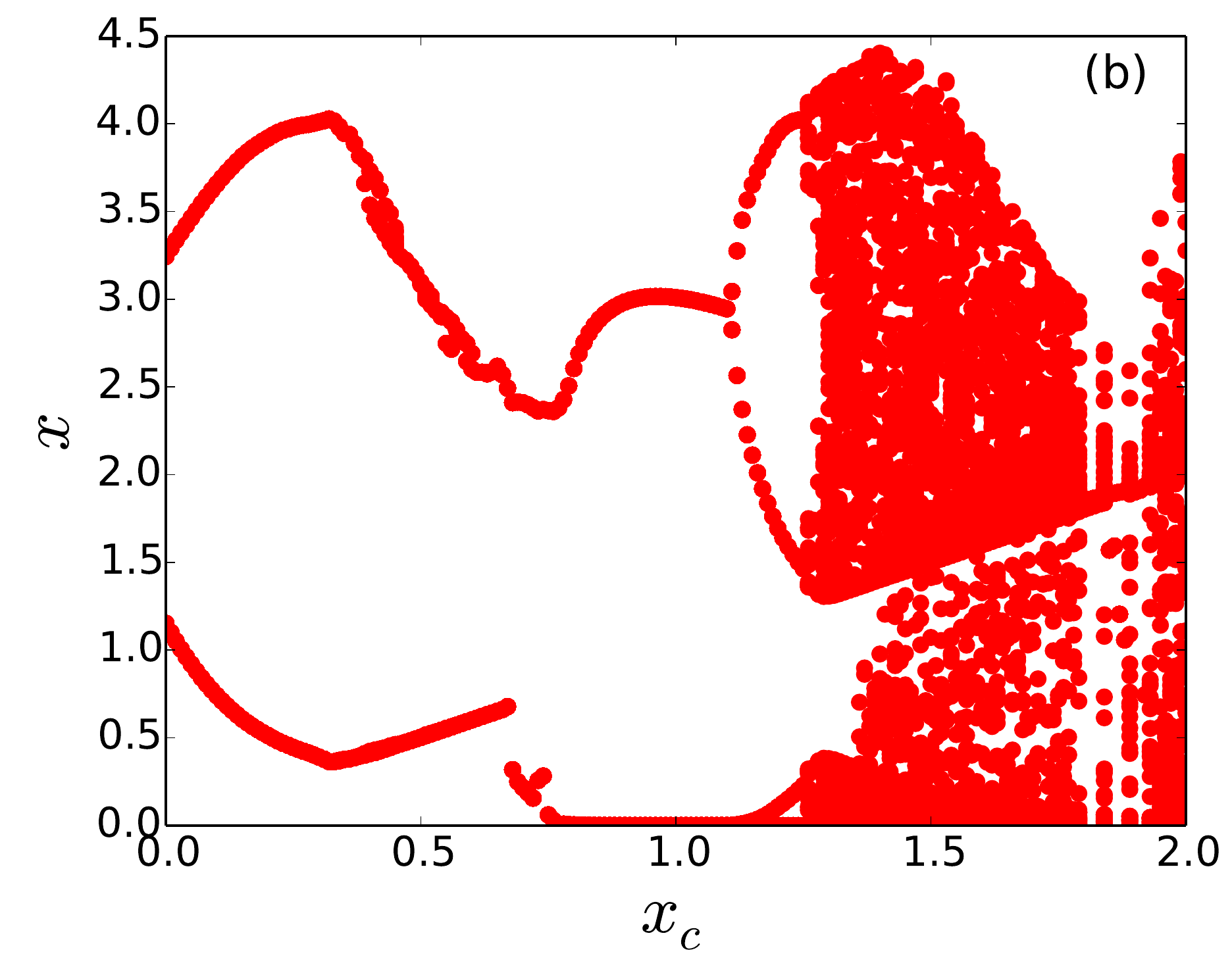}
    \includegraphics[width=\twofigMAT]{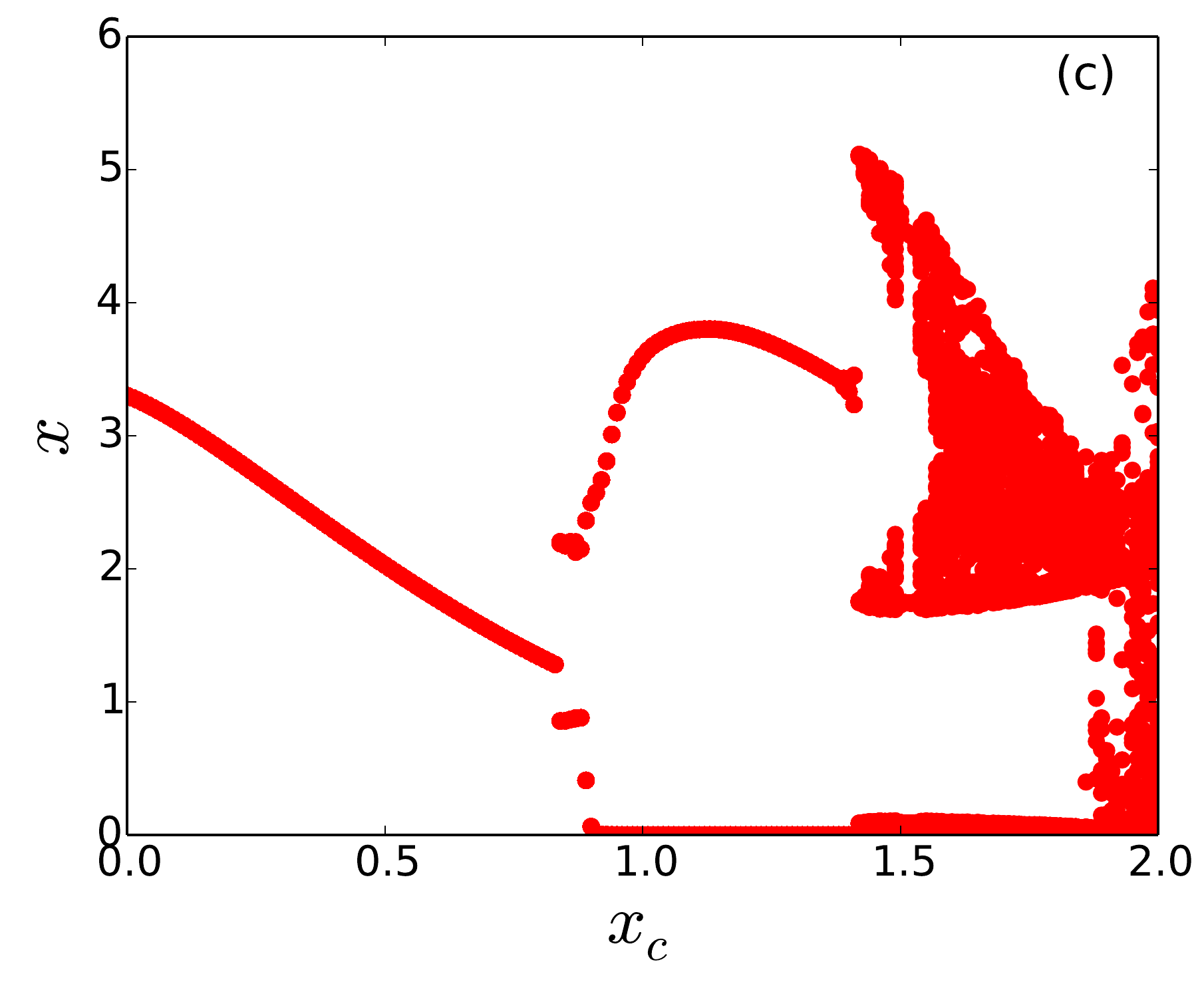}
    \includegraphics[width=\twofigMAT]{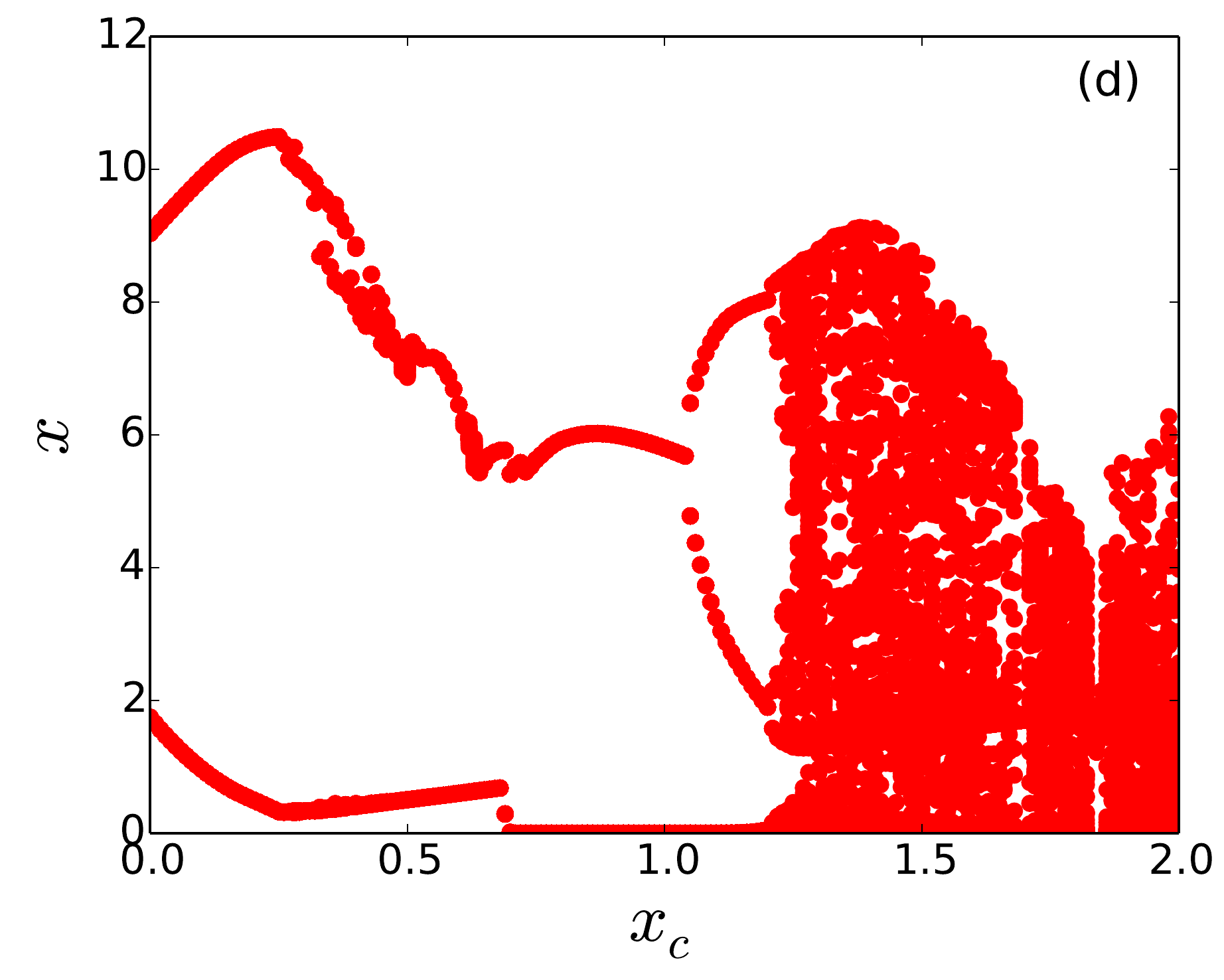}
      \caption{\textbf{Bifurcation diagrams displaying the state of a 
      representative node, with respect to critical threshold $x_c$, 
      in a threshold-coupled Random Scale-Free network of 
      intrinsically chaotic populations.} Here $T_R = 500$ and there 
      is a single open node, with this open node having (a) the 
      highest betweenness centrality, (b) the lowest betweenness 
      centrality, (c) the highest degree and (d) the lowest degree in 
      the network.}\label{BC}
     \end{figure}

		Lastly, we explore the case of networks with very few (typically $1$ or 
		$2$) open nodes, and study the effect of the degree and betweenness 
		centrality \cite{betweeness} of these open nodes on the control to 
		steady states. We observe that when there are very few open nodes, the 
		degree  and betweenness centrality of the open node is important, with 
		the region of control being large when the open node has the high 
		degree/betweenness centrality, and vice versa. This interesting 
		behaviour is clearly seen in the bifurcation diagrams shown in 
		Figs.~\ref{BC}a-d, which demonstrate that the degree and betweeness 
		centrality of the open node has a pronounced influence on control.

		\subsection*{Quantitative Measures of the Efficiency of Chaos Suppression}
		
\begin{figure}[H]
         \centering
         \includegraphics[trim={0.5cm 0 2cm 1cm},clip,width=\onefigMAT]{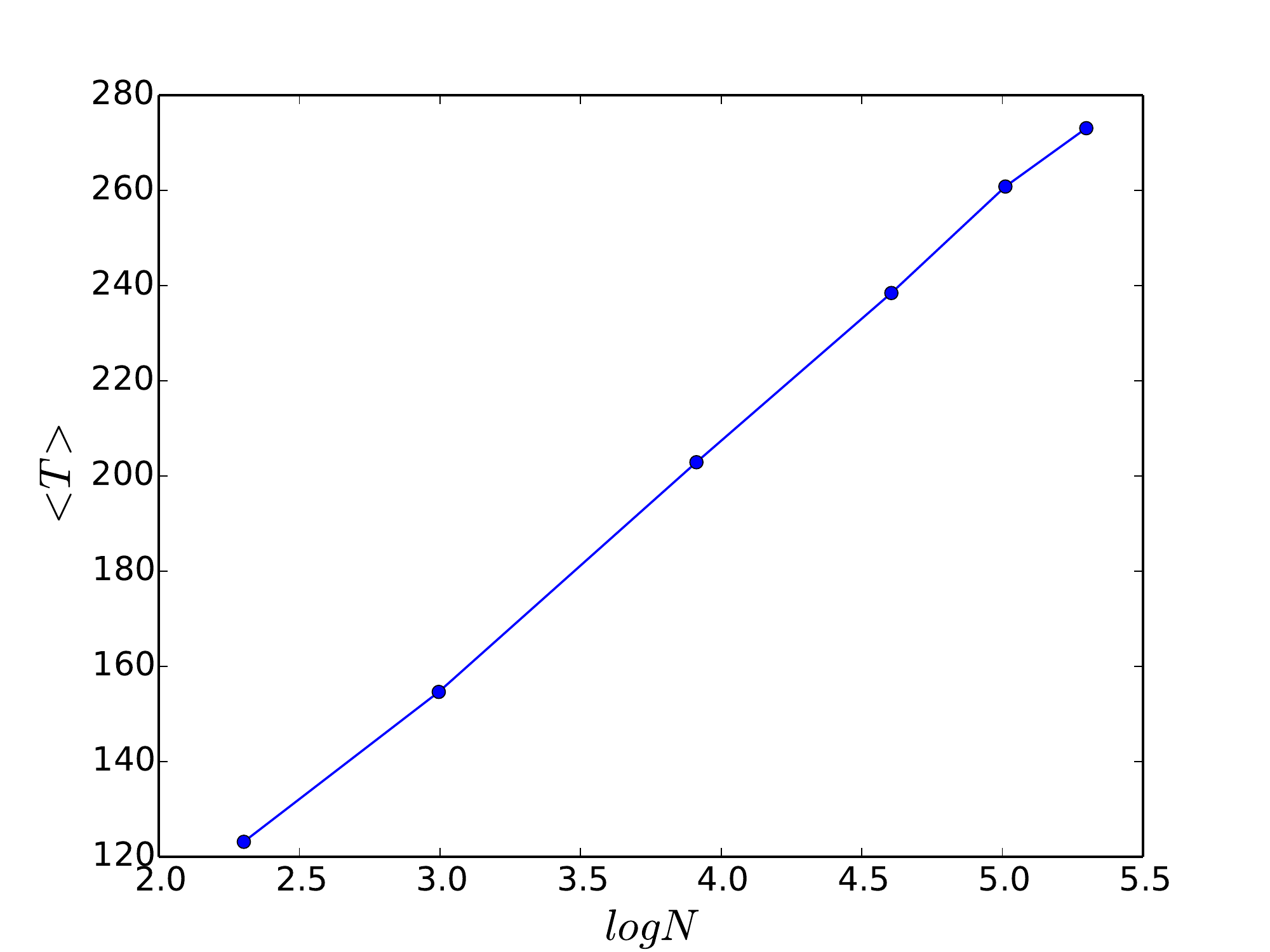}
		\caption{\textbf{Average redistribution time $\langle T
           \rangle$, as a function of the logarithm of the network size $N$.}
            Here $\langle T\rangle$ is defined as the time taken for all nodes in a
           system to be under-critical (i.e. $x_i < x_c, \forall i$),
           averaged over a large sample of random initial states and
           network configurations, the fraction of open nodes in
           the network is $0.2$ and $x_c=0.5$.}\label{NvsRT}
       \end{figure}

\begin{figure}[H]
         \centering
         \includegraphics[trim={0 0 0.2cm 0},clip,width=\onefigMAT]{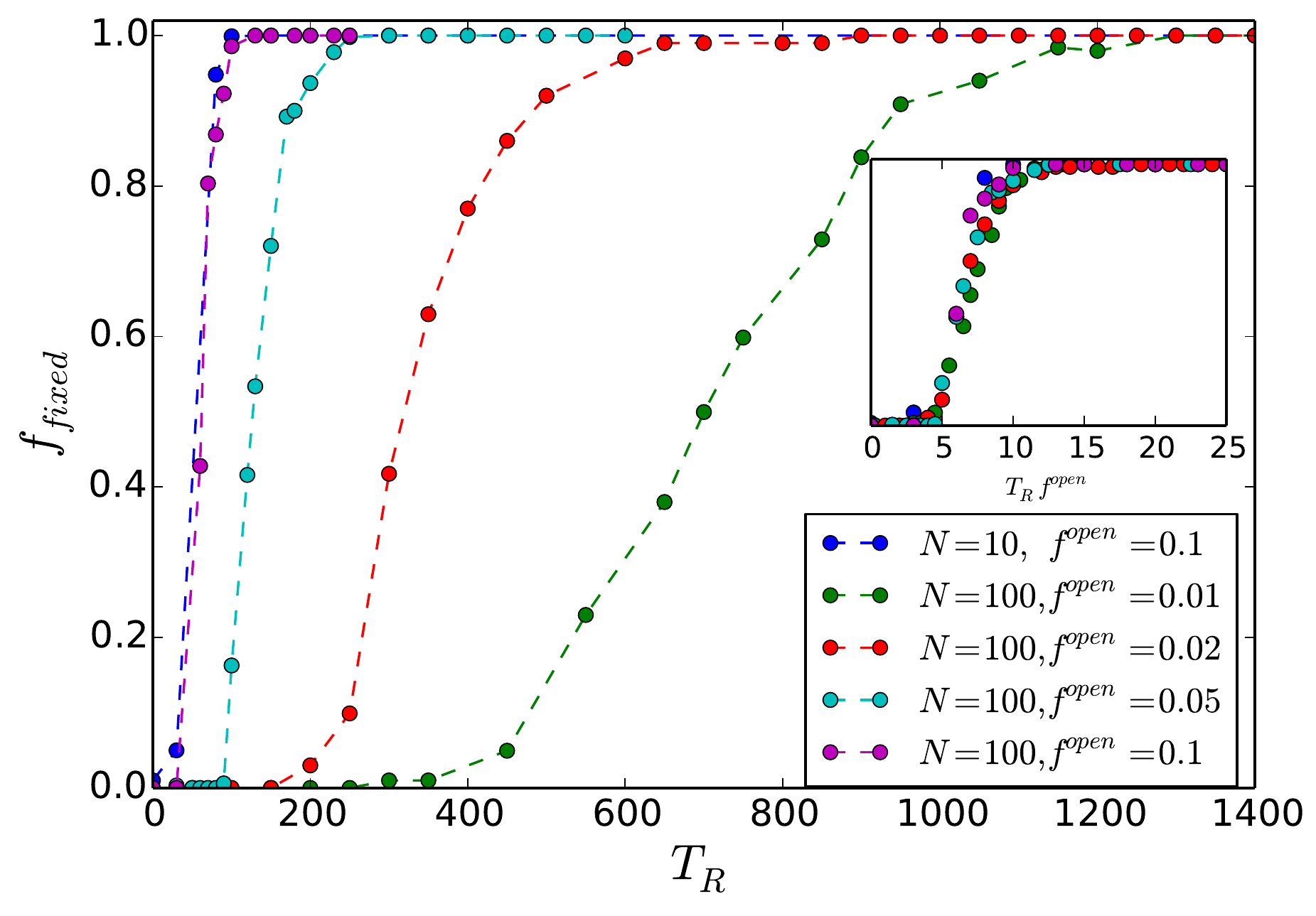}
          \caption{\textbf{Fraction of nodes in the network that go to steady states,
           denoted by $f_{fixed}$, with respect to the redistribution time $T_R$.} 
          Here $f_{fixed}$ is averaged over different network configurations 
          and initial states, $x_c=0.5$ and 
          the fraction of open nodes $f^{open}$  in the network 
          is $0.01, 0.02, 0.05, 0.1$ for 
          $N=100$ (i.e. $1,2, 5, 10$ open nodes in the 
          network respectively) and $0.1$ for $N=10$ 
          (i.e. $1$ open node in the network). {\em Inset}:  
          data collapse indicating the scaling relation
          $f_{fixed} \sim g(T_{R}f^{open})$. 
          }\label{Frac_NodesVsRT}
      \end{figure}

\begin{figure}[H]
         \centering
         \includegraphics[width=\twofigMAT]{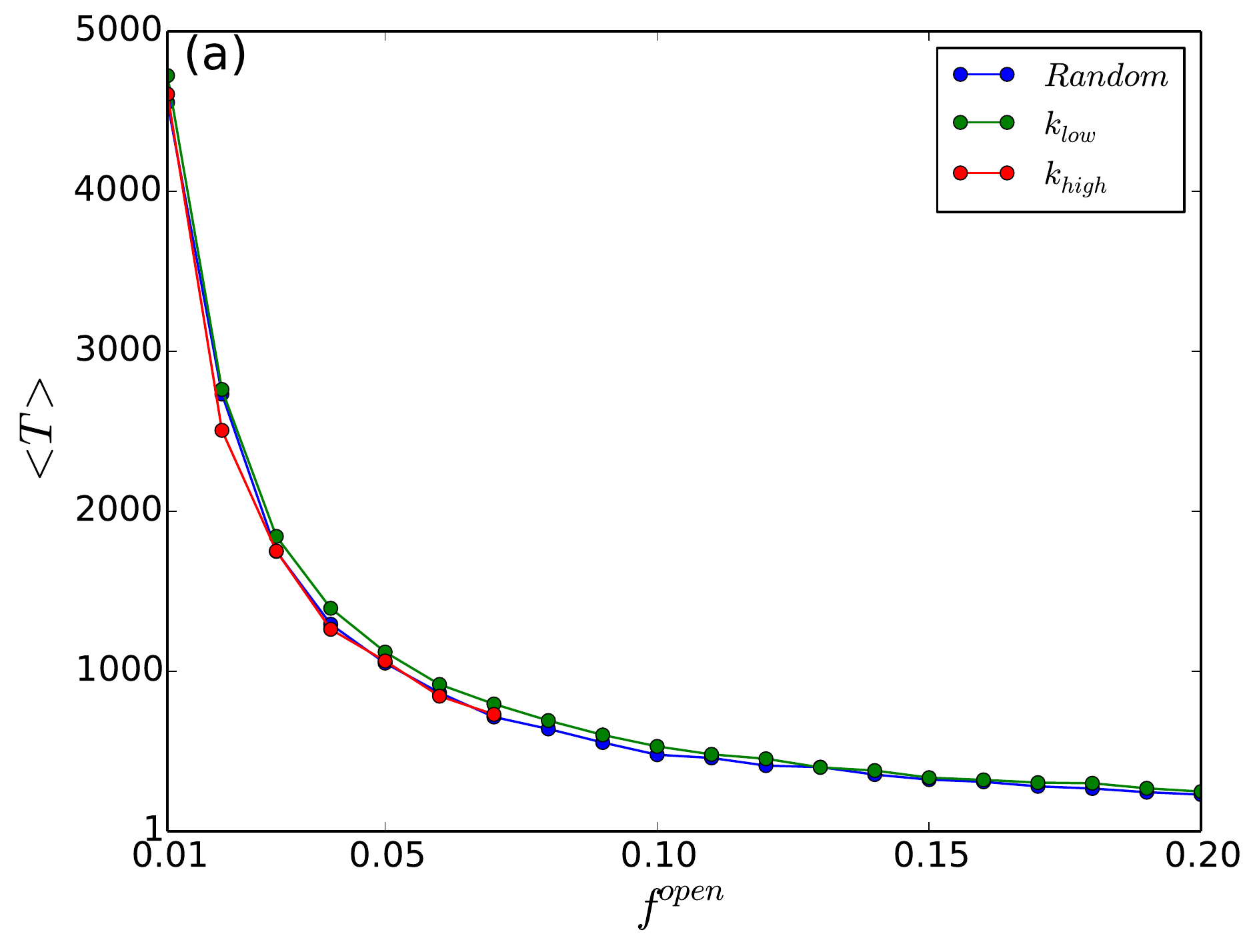}
         \includegraphics[width=\twofigMAT]{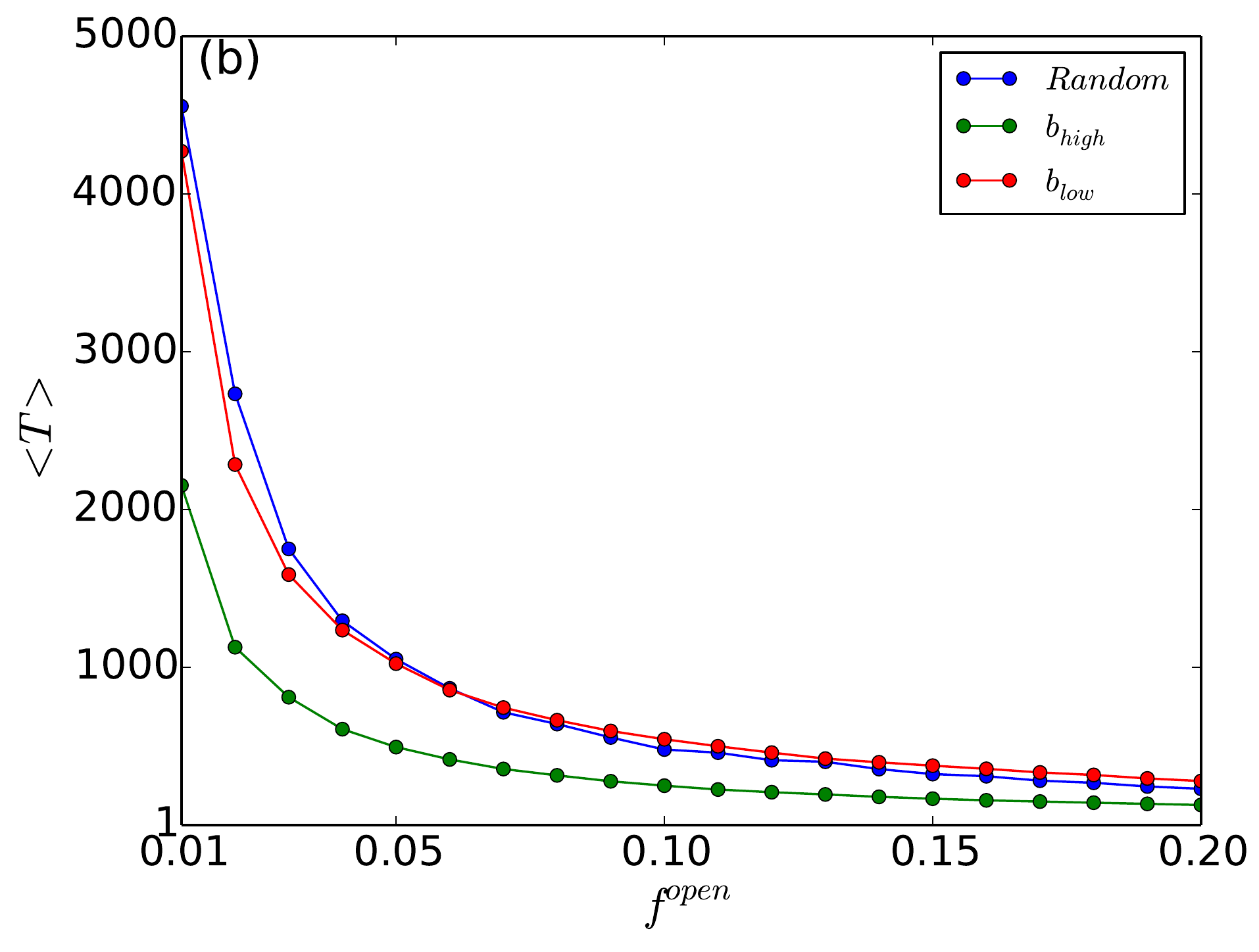}
         \caption{\textbf{Average redistribution time $\langle 
         T \rangle$, as a function of the fraction of open nodes in the network $f^{open}$ }. Here $\langle T \rangle$ is
          defined as the time taken for all nodes in 
         the threshold-coupled Random Scale-Free Network of chaotic 
         populations, to be under-critical (i.e. $x_i < x_c, 
         \forall i$), averaged over a large sample of random 
         initial states and network configurations. There are $100$ chaotic populations 
         connected via threshold-activated transport in a Random 
         Scale-Free network. In panel (a) the case of open nodes 
         chosen in descending order of degree starting from nodes 
         with the highest $k$ (marked as $k_{high}$) and the case 
         of open nodes chosen in ascending order of degree starting 
         from nodes with the lowest $k$ (marked a $k_{low}$), are 
         displayed. In panel (b) the case of open nodes chosen in 
         descending order of betweeness centrality starting from 
         nodes with the highest $b$ (marked as $b_{high}$) and the 
         case of  open nodes chosen in ascending order of 
         betweeness centrality starting from nodes with the lowest 
         $b$ (marked a $b_{low}$), are displayed. In both panels, 
         the case of open nodes chosen at random is also shown for 
         reference. }\label{avg_RT}
         \end{figure}

		We now investigate a couple of quantitative measures that provide
		indicators of the efficiency and robustness of the suppression of
		chaos in the network. The first quantity is the average redistribution
		time $\langle T \rangle$, defined as the time taken for all nodes in a
		system to be under-critical (i.e. $x_i < x_c$ for all $i$), averaged
		over a large sample of random initial states and network
		configurations. So $\langle T \rangle$ provides a measure of the
		efficiency of stabilizing the system, and reflects the rate at which
		the de-stabilizing ``excess'' is transported out of the
		network. Fig.~\ref{NvsRT} shows the dependence of $\langle T \rangle$
		on system size $N$. Clearly, while larger networks need longer
		redistribution times in order to reach steady states, this increase is
		only logarithmic. This is further corroborated by calculating the
		average fraction of nodes in the network that go to steady states with
		respect to the redistribution time $T_R$, for networks of different
		sizes, with varying number of open nodes
		(cf. Fig.~\ref{Frac_NodesVsRT}).  Clearly for small systems, with
		sufficiently high $f^{open}$, very low $T_R$ can lead to stabilization
		of all nodes. Importantly, when the fraction of open nodes is very
		small, the average redistribution time $\langle T \rangle$ depends
		sensitively on the betweenness centrality of the open node, and to a
		lesser extent its degree. Figs.~\ref{avg_RT}a-b present illustrative
		results demonstrating this observation.

\begin{figure}[H]
         \centering
         \includegraphics[trim={0.5cm 0 1.3cm 0.5cm},clip,width=\onefigMAT]{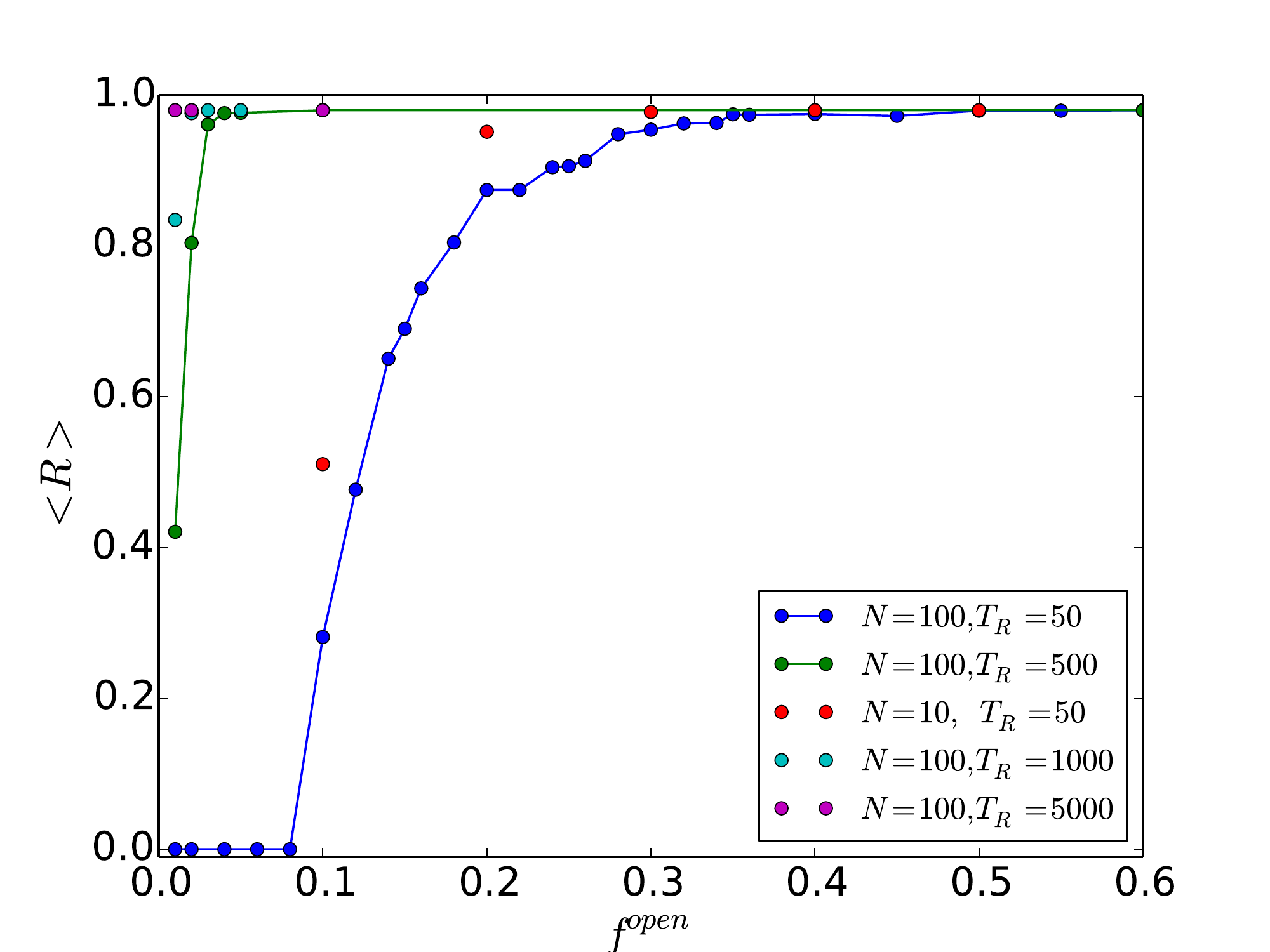}
         \caption{\textbf{Range of threshold values that yield steady states,
          $\langle R \rangle$, as a function of the fraction of 
          open nodes in the network $f^{open}$}. Here  $\langle R \rangle$
          is averaged over different network configurations and initial states and the open 
         nodes are randomly chosen. Results from different redistribution 
         times ($T_R = 50, 500, 1000, 5000$) and system sizes ($N= 10, 
         100$) are shown. }\label{ssrandom}
      \end{figure}

     Next we calculate the {\em range of threshold values yielding steady 
     states}, averaged over a large sample of network configurations and 
     initial states, denoted by $\langle R \rangle$. Larger $\langle R \rangle$ 
     implies that steady states will be obtained in a larger window in $x_c$ 
     space, thereby signalling a more robust control. We have explored the 
     dependence of this quantity on redistribution time $T_R$, and also on the 
     fraction of open nodes in the network, denoted by $f^{open}$. From Fig. 
     \ref{ssrandom} we see that the steady-state window in $x_c$ rapidly 
     converges to $\sim 1$ (namely, the range $0 \le x_c < 1$), as the number 
     of open nodes increases. So the window yielding suppression of chaos is 
     almost independent of the number of open nodes, after a critical fraction 
     of open nodes $f_c^{open}$. We observe that $f_c^{open}$ tends to zero as 
     the redistribution times increases and system size decreases, implying 
     that {\em very few open nodes are necessary in order to lead the network 
     to a steady state}.

\begin{figure}[H]
         \centering
         \includegraphics[width=\onefigMAT]{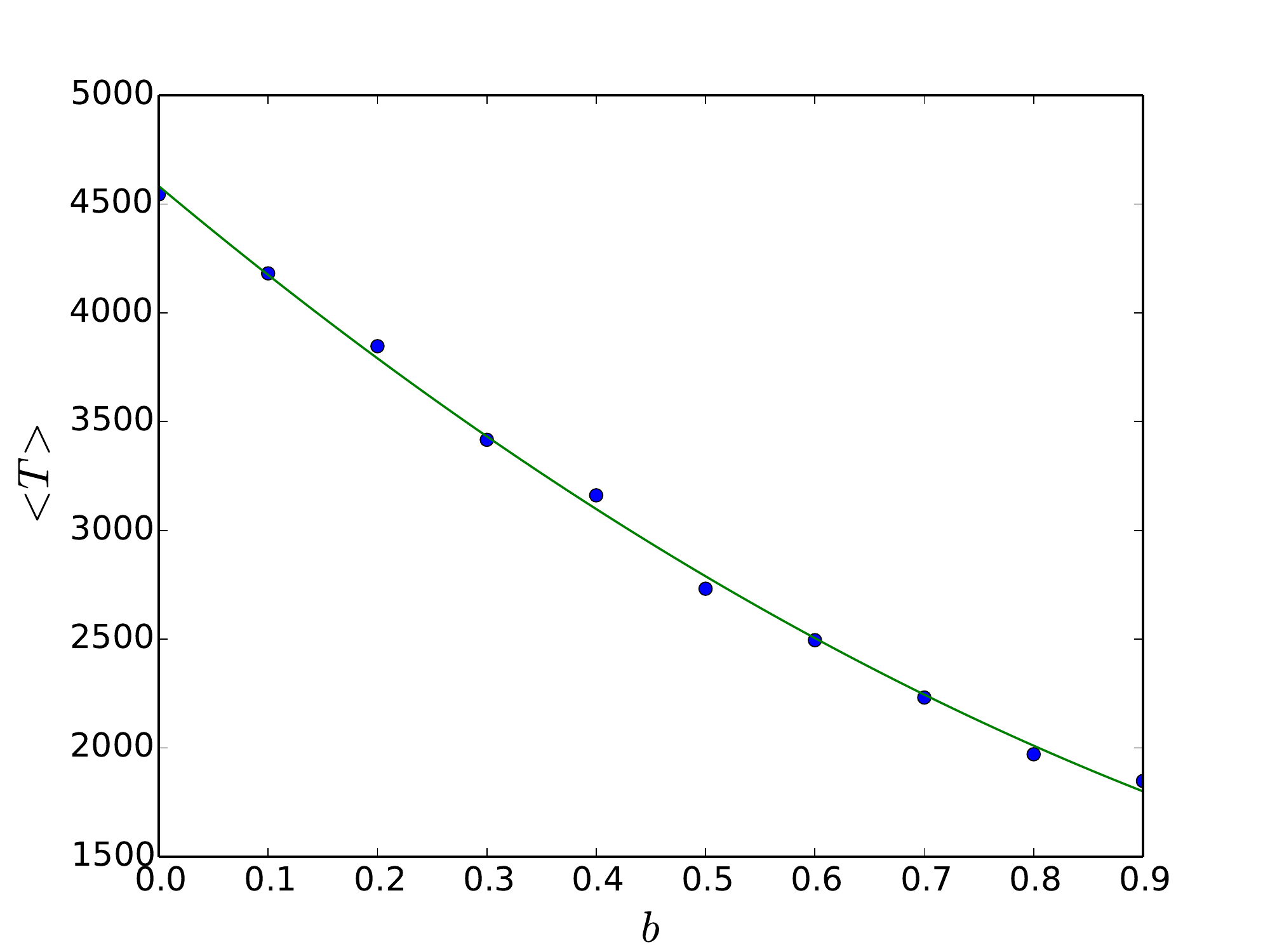}
         \caption{\textbf{Average redistribution time $\langle T \rangle$,
          as a function of the betweeness centrality $b$ of the open node.} Here 
          $\langle T \rangle$ is defined as
          the time taken for all nodes in the threshold-coupled Random Scale-Free Network 
          of chaotic populations, to be under-critical (i.e. $x_i < x_c, \forall i$), 
          averaged over a large sample of random initial states and network configurations,
           in a network with a single open node. The solid curve shows the best quadratic polynomial fit.
            }\label{btcvsRT}
     \end{figure}

\begin{figure}[H]
         \centering
         \includegraphics[width=\twofigMAT]{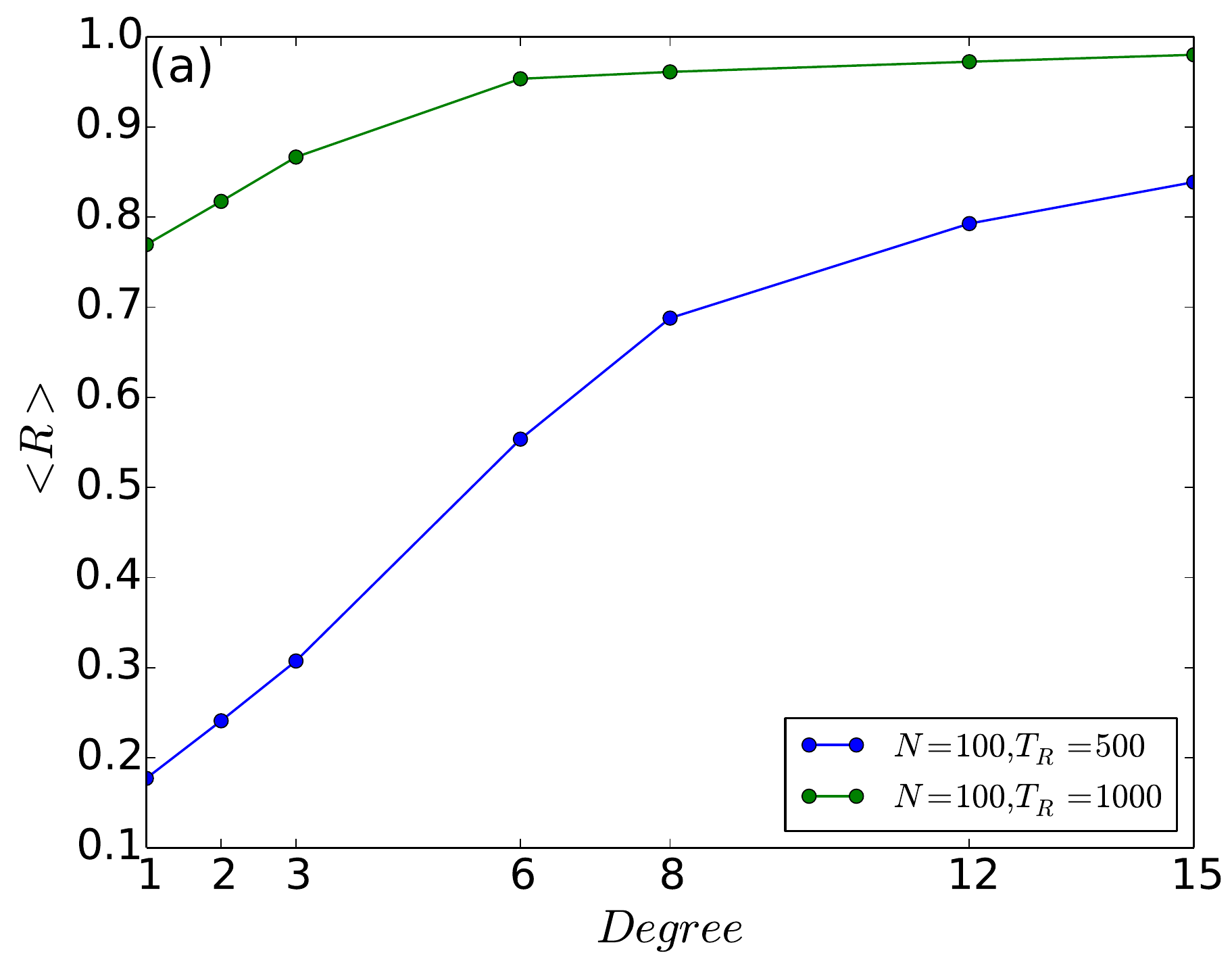}
         \includegraphics[width=\twofigMAT]{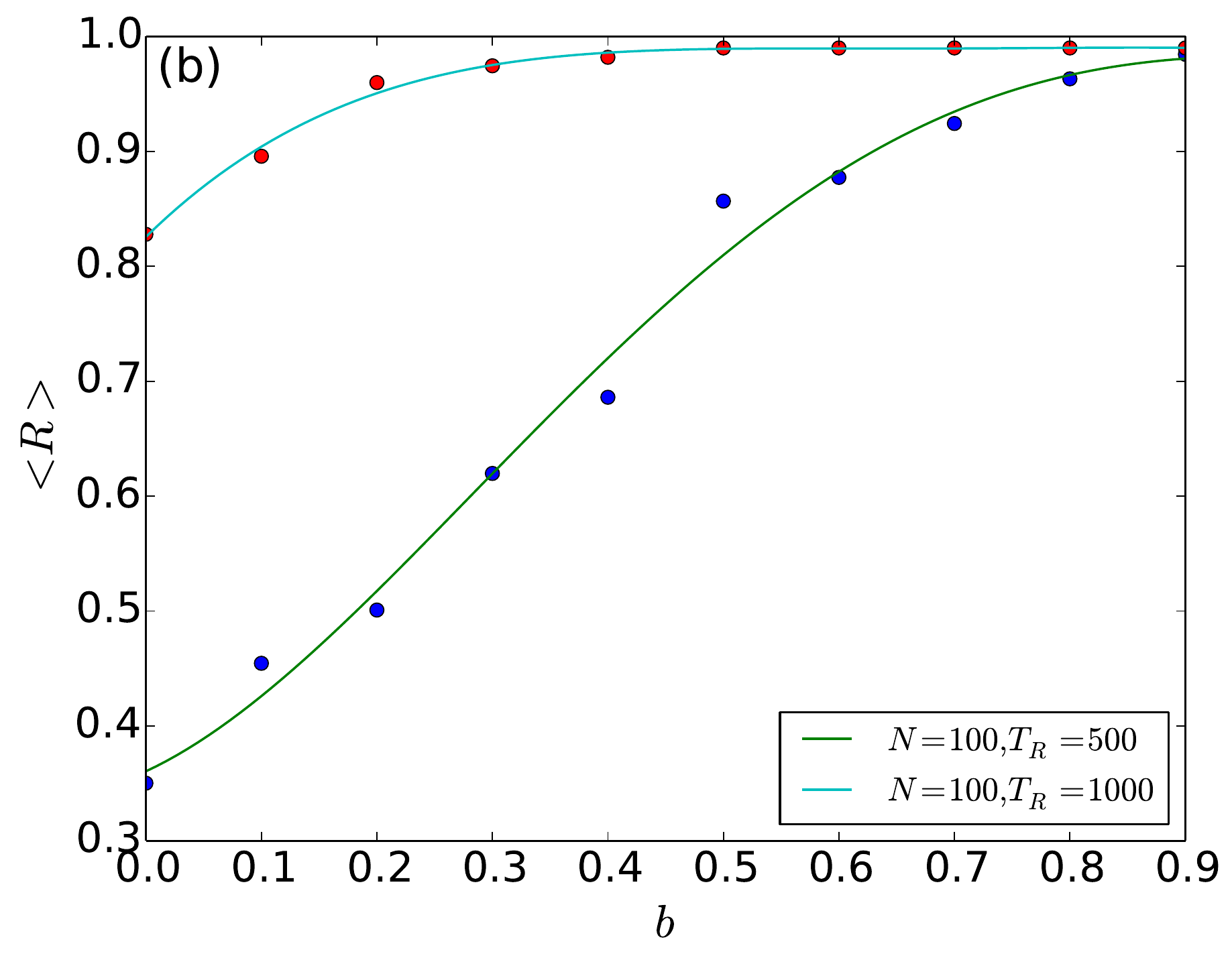}
         \caption{\textbf{Range of threshold values that yield steady states
         $\langle R \rangle$, as a function of the (a) degree $k$, and (b)
           betweeness centrality $b$, of the open node.} Here $\langle R \rangle$ is averaged
         over different network configurations and initial states, 
        in a network with a single open node, (with the solid curve showing the best 4th order polynomial fit). 
		% The solid curves show the best quadratic polynomial fits: $f(b) = p b^4 + q b^3 + r b^2 + s b + t$, where $p = 0.361, q = 0.466, r = 2.206, s = -3.314, t = 1.265$ for $R_T=500$ and  $p =  0.826, q = 0.983, r = -2.185, s = 2.130, t = -0.767$ for $R_T=1000$.
				}\label{RangeVsDegR}
	  \end{figure}

   Lastly we explore the scenario of very few open nodes ($f^{open} <<
   f_c^{open}$) in greater depth, through the quantitative measures
   $\langle R \rangle$ and $\langle T \rangle$. In particular, we
   investigate the limiting case of a {\em single} open node. Our
   attempt will be to understand the influence of the degree $k$ and
   betweeness centrality $b$ of the open node on the capacity to
   suppress chaos. We have already observed the significant effect of
   the betweeness centrality of the open node on the efficiency of
   control to steady states through bifurcation diagrams in
   Fig.~\ref{BC}. This is now further corroborated quantitatively by
   the dependence of $\langle R \rangle$ and $\langle T \rangle$,
   displayed in Figs.~\ref{btcvsRT} and \ref{RangeVsDegR}(b). The
   effect of the degree of the open node is less pronounced, though it
   also does have a discernable effect on the suppression of chaos. As
   evident from Fig.~\ref{RangeVsDegR}(a), when the open node has a
   higher degree, it has a higher $\langle R \rangle$, indicating that
   open nodes with higher degree yield larger steady state windows.

	\section*{Conclusions}	

	We have explored Random Scale-Free networks of populations under 
	threshold-activated transport. Namely we have a system comprising of many 
	spatially distributed sub-populations connected by migrations triggered by 
	excess population density in a patch. We have simulated this 
	threshold-coupled Random Scale-Free network of populations, under varying 
	threshold levels $x_c$.  We considered networks with varying number of open 
	nodes, namely systems that have different nodes/sites open to the 
	environment from where the excess population can migrate out of the system. 
	Further, we have studied a range of redistribution times $T_R$, capturing 
	different timescales of migration vis-a-vis population change. 

Our first important observation is as follows: when redistribution time $T_R$ 
is large and the critical threshold $x_c$ is small ($0 \le x_c < 1$), we have 
very efficient control of networks of chaotic populations to steady states. 
This suppression of chaos and quick evolution to a stable steady states occurs 
irrespective of the number of open nodes. Further, for threshold values beyond 
the window of control to fixed states, one obtains cycles of period $2$. Namely 
for threshold levels $1 < x_c < 2$ the populations evolve in regular cycles, 
where low population densities alternate with a high population densities. This 
behaviour is reminiscent of field experiments \cite{self-perpetuating} that 
show the existence of alternating states. We offer an underlying reason for 
this behaviour through the analysis of a single threshold-limited map.

For small redistribution time $T_R$, the system does not get enough time to 
relax to under-critical states and so perfect control to steady states may not 
be achieved. Importantly, now the number of open nodes is crucial to chaos 
suppression. We clearly demonstrate that when there are enough open nodes, the 
network relaxes to the steady state even for low redistribution times. So more 
open nodes yields better control of the intrinsic chaos of the nodal population 
dynamics to fixed populations. We corroborate all qualitative observations by 
quantitative measures such as average redistribution time, defined as the time 
taken for all nodes in a system to be under-critical, and the range of 
threshold values yielding steady states.

We also explored the case of networks with very few (typically $1$ or $2$) open 
nodes in detail, in order to gauge the effect of the degree and betweenness 
centrality of these open nodes on the control to steady states. We observed 
that the degree of the open node does not have significant influence on chaos 
suppression. However, betweenness centrality of the open node is important, 
with the region of control being large when the open node has the high 
betweenness centrality, and vice versa.

In summary, threshold-activated transport yields a very potent coupling form in 
a network of populations, leading to robust suppression of the intrinsic chaos 
of the nodal populations on to regular steady states or periodic cycles. So 
this suggests a mechanism by which chaotic populations can be stabilized 
rapidly through migrations or dispersals triggered by excess population density 
in a patch.

     \section*{Appendix : Analysis of a single population patch under 
     threshold-activated transport}

     We will now analyze the dynamics of a single Ricker map, modelling a 
     single population patch, under threshold-activated transport. Specifically 
     then we have the following scenario: in the dynamical evolution of the 
     system, if the updated state exceeds a critical threshold $x_c$, it 
     transports the excess out of the system and``re-sets'' to level $x_c$. So 
     the effective map of the dynamics is:
     \begin{eqnarray}
     x_{n+1}=f(x_n) \ \ \ {\rm if } \ \ f(x_n) < x_c \\ \nonumber
     x_{n+1}=x_c \ \ \ {\rm if } \ \ f(x_n) \ge x_c \\
     \end{eqnarray}

\begin{figure}[!h]
				\centering 
               \includegraphics[width=\twofigMAT]{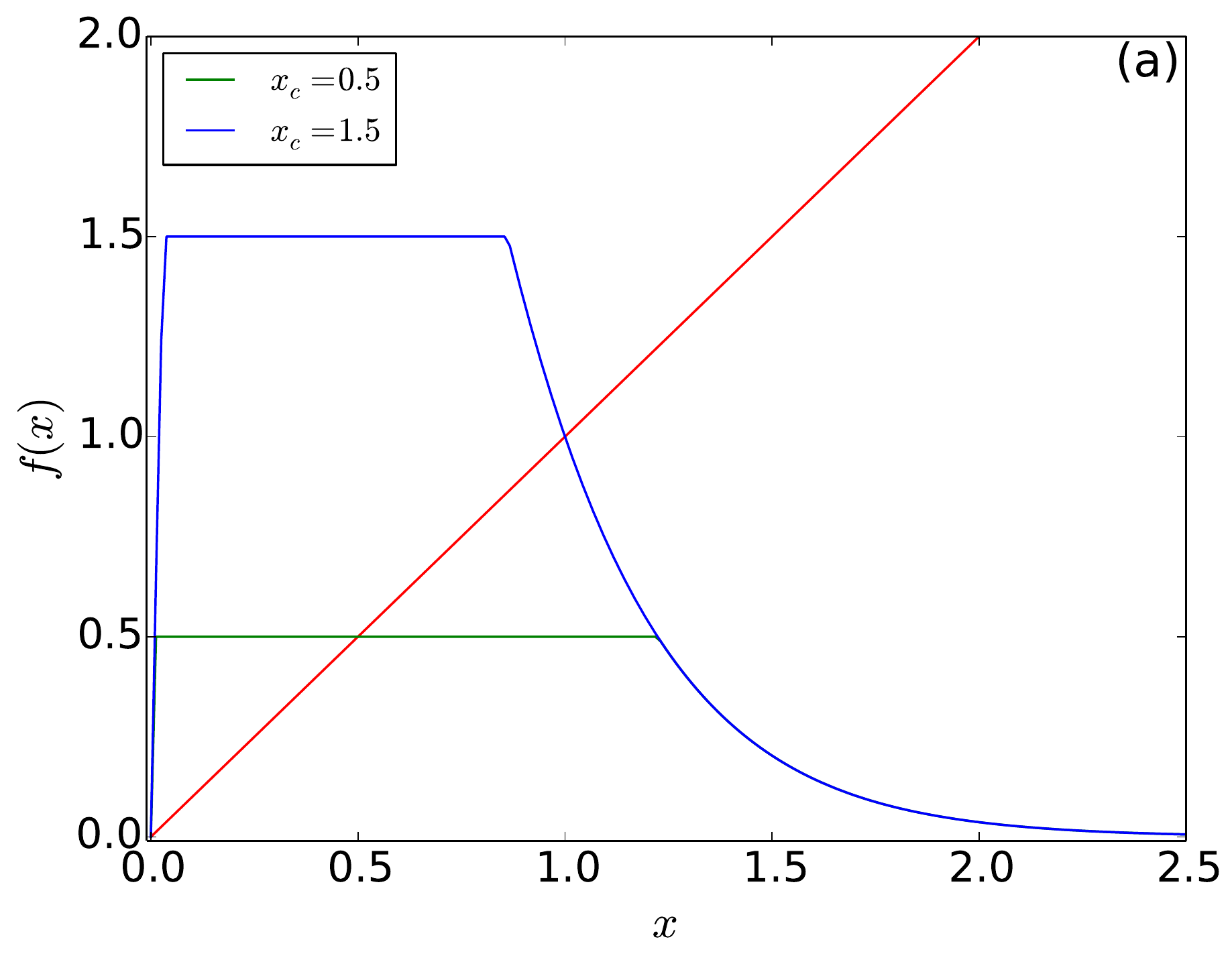}
                \includegraphics[width=\twofigMAT]{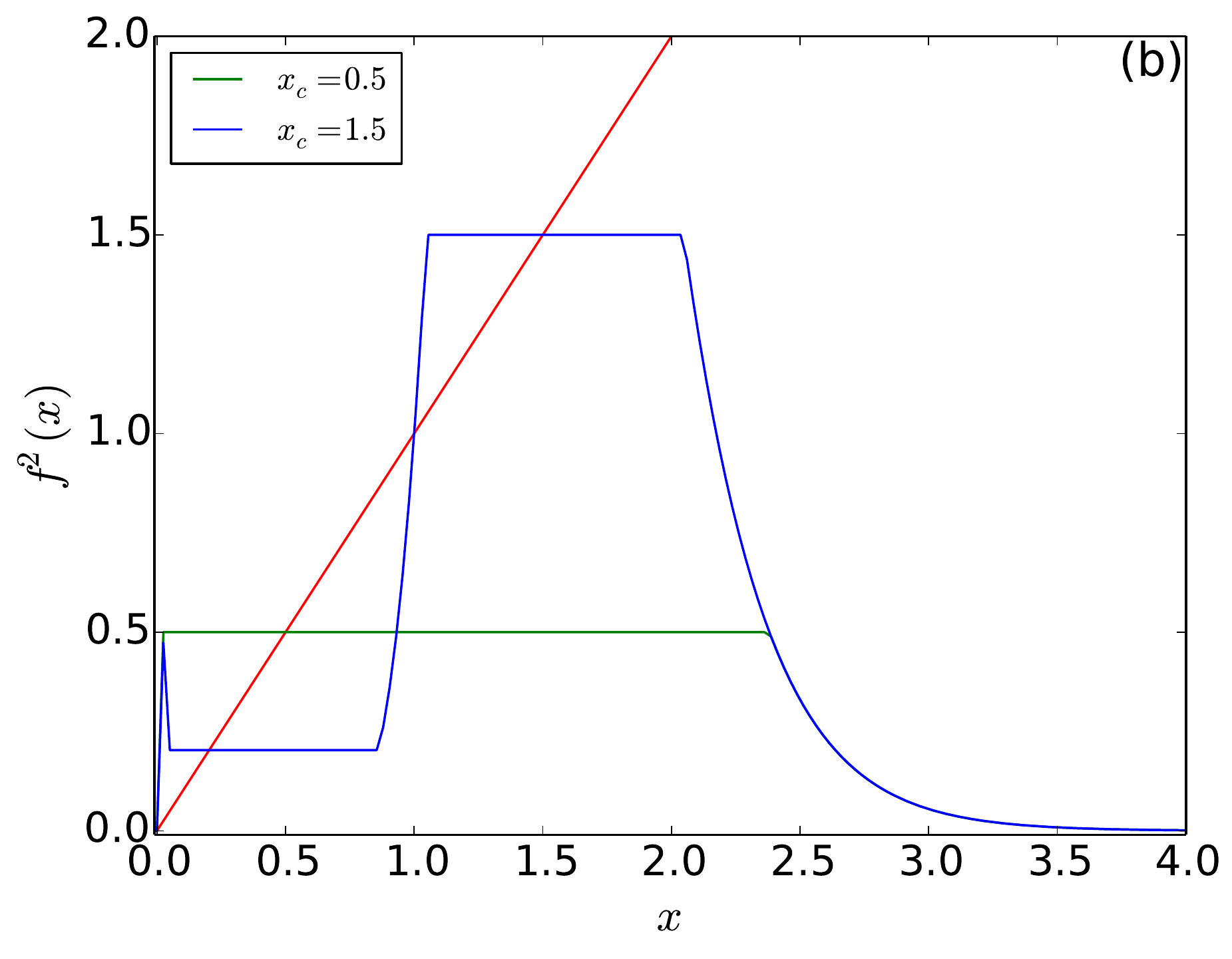}
				\caption{\textbf{(a) $f(x)$ vs. $x$ and (b) $f^2(x)$ vs. $x$, for the 
				effective threshold-controlled Ricker map ($r=4$), for critical 
				threshold levels: $x_{c}=0.5$ (green) and $x_c = 1.5$ (blue).}
				The fixed point solution occurs at the intersection of the 
				$f(x)$ curve and the $45^0$ line, and is stable if the 
				intersection occurs at the ``flat top'', namely where 
				$f^{\prime} (x) = 0$.}
				\label{controlfn}
             \end{figure}

     This is effectively a ``beheaded'' or ``flat-top'' map, with the curve 
     lying above $x_{n+1}>x_c$ in the usual Ricker map being ``sliced'' to 
     $x_c$ (cf. Fig.~\ref{controlfn}a). The level at which the map is chopped 
     off depends on the threshold $x_c$. The fixed point solution $x^{\star}$ 
     occurs at the intersection of this $f(x)$ curve and the $45^0$ line, 
     namely $x^{\star} = x_c$. Remarkably, this fixed point is {\em 
     super-stable} if the intersection occurs at the ``flat top'', since 
     $f^{\prime} (x^{\star}) = 0$ there.

Clearly, as the threshold increases the intersection of the effective map and 
the $45^0$ line is no longer located at the ``flat-top''. This is clear for the 
effective maps for $x_c=0.5$ vis-a-vis that for $x_c = 1.5$ in 
Fig.~\ref{controlfn}a. So $x^{\star}$ for sufficiently high $x_c$ will no 
longer be stable (eg. $x_c = 1.5$ will not yield a stable fixed point). So we 
go on to inspect the second iterate of the effective map, in order to ascertain 
if a stable period-$2$ cycle is obtained (cf. Fig.~\ref{controlfn}b). Now the 
period-$2$ cycle solutions occur at the intersection of the $f^2(x)$ curve and 
the $45^0$ line, and again this cycle is stable if and only if the intersection 
occurs at the ``flat top'', namely where $f^{\prime} (x) = 0$. In the 
illustrative example displayed in Fig.~\ref{controlfn}b it is clear that for 
$x_c=0.5$, where the fixed point is super-stable, the period-$2$ is also 
naturally super-stable. Interestingly now, for $x_c=1.5$, which had an unstable 
fixed point solution, the period-$2$ solution is super-stable. So higher $x_c$ 
also controls the intrinsic chaos. However, instead of a stable steady state, 
it yields stable periodic behaviour.

Alternately, one can understand the emergence of stable cycles under threshold 
control as follows: The ergodicity of the system ensures that the system will 
explore the available phase space fully, and the state variable is thus 
guaranteed to exceed threshold at some point in time. So one can analyse the 
dynamics of the effective map starting with the initial state at $x_c$. Now 
starting from $x_c$ the dynamics will run as in the usual Ricker population map 
until $x_{n+1}>x_c$, at which point it is re-set back to $x_c$ and the cycle 
starts again. So once it exceeds the critical value it is trapped immediately 
in a stable cycle whose periodicity is determined by the value of the 
threshold. Further, this allows us to exactly obtain the values of threshold 
$x_c$ that yield stable fixed points $x^*$ (namely period-$1$). This is simply 
the range of $x_c$ for which the first iterate of the Ricker map lies above 
$x_c$. In this range $f (x_c) > x_c$. So starting from an initial state $x_c$, 
we will be updated in the next iterate to a state greater than $x_c$, leading 
to the transport of the excess $f(x)-x_c$ out of the system and the 
``relaxation'' of the system to $x_c$.

\begin{figure}[H]
			\centering 
			\includegraphics[width=\onefigMAT]{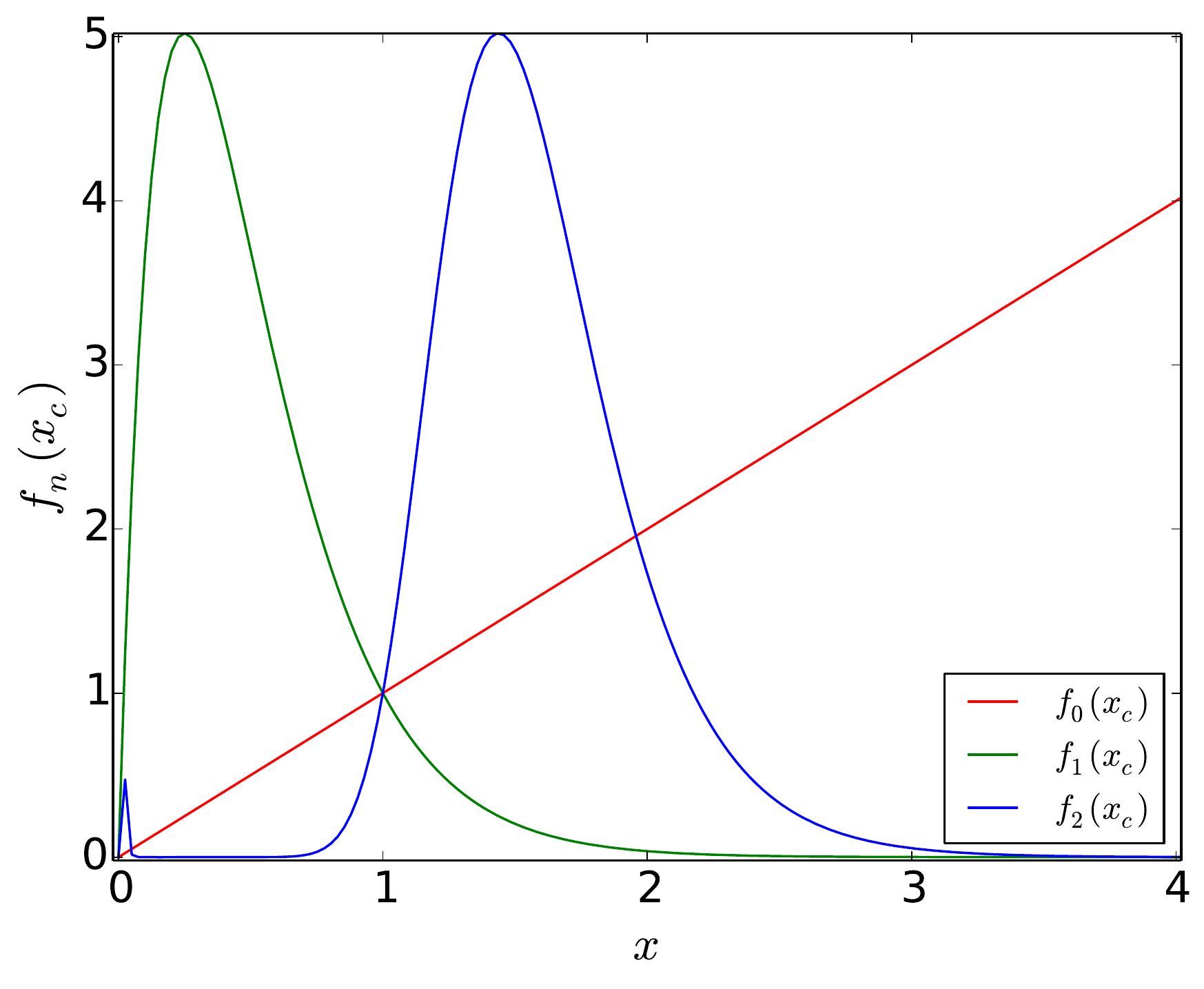}
					\caption{\textbf{Plot of $f_{n} (x_{c})$ vs $ x_{c}$.} Here $n=0, 1, 
					2$, where $f_{n} (x_{c})$ is the $n^{th}$ iterate starting 
					from initial condition $x=x_{c}$ of the Ricker map with 
					$r=4$: $f_{0} (x_{c})$ (red), $f_{1} (x_{c})$ (green) and 
					$f_{2} (x_{c})$ (blue).}		
			  		\label{fn}
		\end{figure}

              The curves $f_{n} (x_{c})$ as a function of threshold $x_{c}$ are 
              displayed in Fig.~\ref{fn}. For $n=0$, $f_{0}(x_{c})= x_{c}$; for 
              $n=1$, $f_{0}(x_{c})=  x_{c}\exp(r(1- x_{c}))$, and in general 
              $f_{n}(x_{c}) = f \circ f_{n-1}(x_{c}) = f \circ f \circ  \dots 
              f(x_{c})$. From the figure it can be clearly seen that in the 
              range of $x_c \in [0:1]$, $f(x_c) > x_c$. So if the threshold is 
              in this range, the system will evolve quickly to a steady state 
              at $x^*=x_c$, and transport the excess, namely $f(x_c)-x_c$, out 
              of the system after every update of the population in the patch. 

             Similarly, it can be seen that $f_2 (x_c) = f (f (x_c))$ is larger 
             than $x_c$ (while $f (x_c) < x_c$) in the range of threshold $x_c 
             \in (1,2]$. So in this range of threshold, we obtain a stable 
             period $2$ cycle. Namely, the population at $x_c$ evolves to 
             $f(x_c)< x_c$ which then evolves to $f^2 (x_c)$. Since  $f^2 
             (x_c)> x_c$, it is mapped back to $x_c$. Hence a cycle of period 
             $2$ arises, with the values of the two points in the cycle being  
             $x_c$ and $f (x_c)$. It can be seen from Fig.~\ref{fn} that this 
             range is from $x_c \sim 1$ to $x_c \sim 2$. This also corroborates 
             the analysis using effective ``flat-top'' maps (cf. 
             Fig.~\ref{controlfn}).

\begin{figure}[!h]
		\centering 
		\includegraphics[width=\onefigMAT]{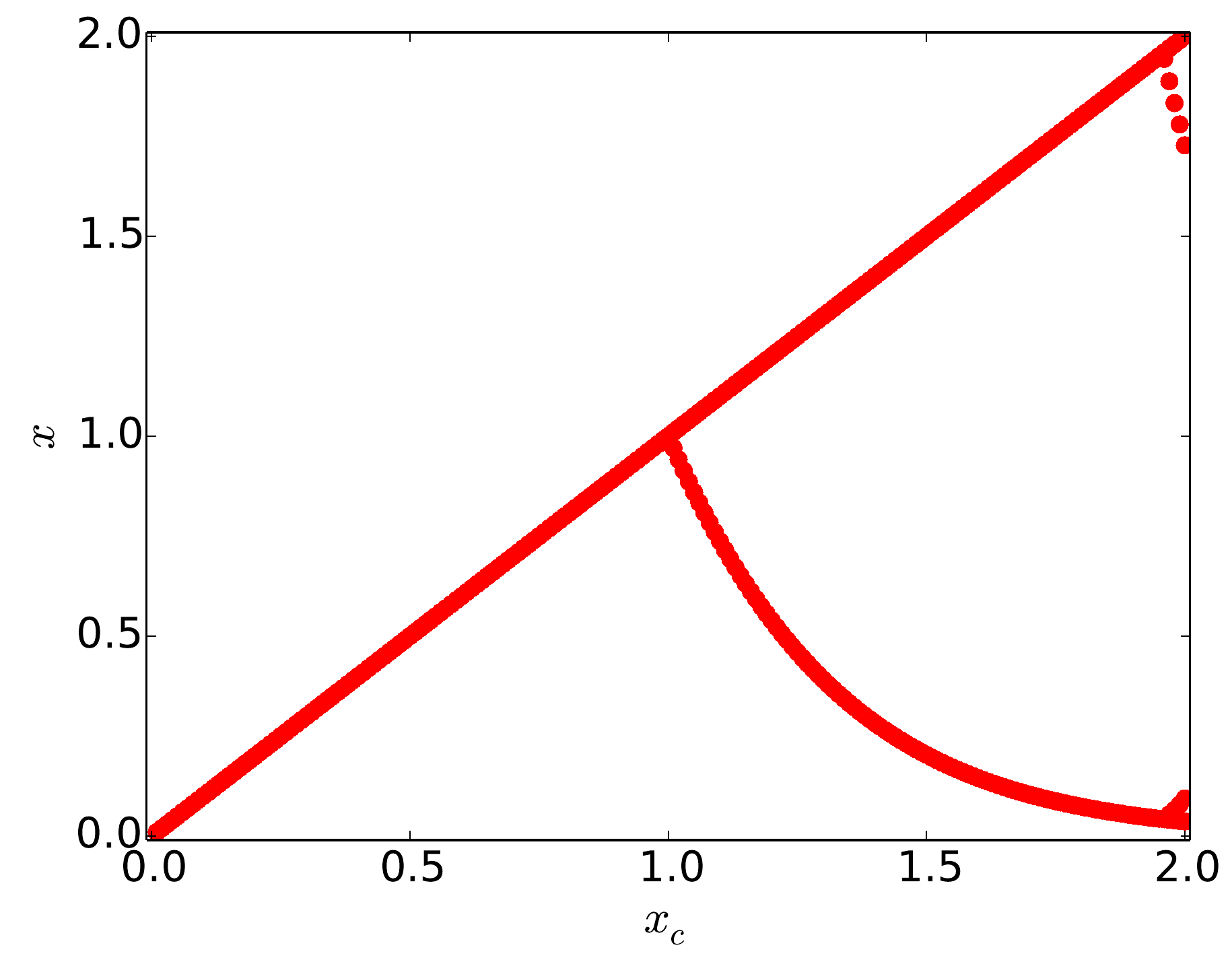}
		\caption{\textbf{Bifurcation diagram of the threshold-controlled Ricker map 
		(namely, the ``flat-top'' map), with respect to threshold level 
		$x_c$.}} \label{bif}
	\end{figure}

	When there is enough time to relax between chaotic updates (namely $T_R$ is 
	large and/or the number of open nodes is sufficiently high), the collective 
	excess of the network is transported out of the system. This implies that the 
	individual nodes behave essentially like the ``flat-top'' map analysed here. 
	This explains why the range of threshold values yielding fixed points and 
	period-$2$ cycles obtained in networks of threshold-coupled chaotic systems 
	(cf. Fig.~\ref{withedgeRandomSFm1highT_R}) matches so well with that obtained 
	here  (cf. Fig.~\ref{bif}). 
	
\newpage
	\begin{thebibliography}{20}
	
	\bibitem{control} A few representative examples: A. Garfinkel, M. Spano, W. 
	Ditto, J. Weiss, {\em Science} {\bf 257} (1992) 1230; K. Hall et al, {\em 
	Phys. Rev. Letts.} {\bf 78}{4518}{1997}.

	\bibitem{adaptive} B. A. Huberman and E. Lumer, {\em IEEE Trans. Circuits 
	Syst.} {\bf 37} (1990) 547; S. Sinha, R. Ramaswamy and J. Subba Rao,  {\em 
	Physica} {\bf D 43} (1990) 118.

\bibitem{ogy} E. Ott, C. Grebogi, and J.A. Yorke, {\em Phys. Rev. Lett.} {\bf 
64}, 2837 (1990).

	\bibitem{glass} L. Glass and W. Zheng, {\em Int. J. Bif. and Chaos}, {\bf 
	4} (1994) 1061.

\bibitem{pattern_control} S. Sinha and N. Gupte, {\em Phys. Rev. E}, {\bf 58} 
(1998) R5221.

\bibitem{motter} S.P. Cornelius, W.L. Kath, and A.E. Motter, {\em Nature 
communications} {\bf 4} (2013).
	
  	\bibitem{metapopulation_review} I Hanski, {\em Nature}, {\bf 396}, 41 
  	(1998).
  	
  	\bibitem{meta2} C.M. Taylor and R.J. Hall, {\em Biology Letters}, {\bf 8}, 
  	477 (2012).
 
%	\bibitem{levin1}  R Levins, {\em Extinction} in {\em Some mathematical 
%questions in biology} (ed. M. Gerstenhaber), pp. 77-107. Providence, RI: The 
%American Mathematical Society (1970).
	
	\bibitem{levin2} R Levins, Bulletin of the ESA, {\bf 15}, 237 (1969).
 
%	\bibitem{dispersal1} B Kerr, M A Riley, M W Feldman \& B J M Bohannan, {\em 
%Nature}, {\bf 418}, 171 (2002).
	
\bibitem{colizza} 
V. Colizza, R. Pastor-Satorras and A Vespignani, {\em Nature Physics} {\bf 3}, 
276-282 (2007).
%Reaction–diffusion processes and metapopulation models in heterogeneous 
%networks

	\bibitem{dispersal2} B Kerr, C Neuhauser, B J M Bohannan \& A M Dean, {\em 
	Nature}, {\bf 442}, 75 (2006).
	
%	\bibitem{dispersal3} S Dey \& A Joshi, {\em Science}, {\bf 312}, 434 (2006).
	
	\bibitem{dispersal4}
	JH Brown \& A Kodric-Brown, {\em Ecology}, {\bf 58}, 445 (1977).
 
	\bibitem{dispersal5} T Reichenbach, M Mobilia \& E Frey, {\em Nature}, {\bf 
	448}, 1046 (2007).
	
	\bibitem{dispersal6} B Cazelles, S Bottani \& L Stone, {\em Proc. R. Soc. 
	B}, {\bf 268}, 2595 (2001). 
		
%	\bibitem{May72} R. May, Nature, {\bf 238}, 413-414(1972). 
	
%	\bibitem{general} B. Blasius, A. Huppert, L. Stone, {\em Nature} {\bf 399} 
%(1999), 354; T Gross, B Blasius, {\em J. of R. Soc. Interface} {\bf 5} (2008), 
%259; M. Baurmann, T. Gross, U. Feudel, {\em J. of Theor. Bio.} {\bf 245} 
%(2007) 
%220.
	
	%\bibitem{stefano1} G Jacopo, G Barab{\'a}s, and S Allesina, PLoS Comput. 
	%Biol., \textbf{11.5} (2015); S Allesina and S Tang, Population Ecology, 
	%\textbf{57.1}, 63-75 (2015).
	
%	\bibitem{stability1} A. R. Ives \&  S. R. Carpenter, Science, {\bf 317} 
%58-62(2007).
 
%	 \bibitem{stability3} A. Mougi \&  M. Kondoh, Science, {\bf 337}, 
%349-351(2012).
 
%	 \bibitem{sinha} S. Sinha and S. Sinha, Phys. Rev. E, \textbf{71}, (2005) 
%020902; S. Sinha and S. Sinha, Phys. Rev. E, \textbf{74}, (2006) 066117.
	
\bibitem{ssprl} S. Sinha and D. Biswas, {\em Phys. Rev. Letts.} {\bf 71} (1993) 
2010.
	\bibitem{sspre} S. Sinha, {\em Phys. Rev. E}, {\bf 49} (1994) 4832; {\em 
	Phys. Letts. A,} {\bf 199} (1995) 365; {\em Int. Jour. Mod. Phys. B} {\bf 
	9} (1995) 875.  

\bibitem{kazu} T. Morie, D. Atuti, K. Ifuku, Y. Horio, K. Aihara
%A CMOS nonlinear-map circuit array for threshold-coupled chaotic maps using 
%pulse-modulation approach.
(2011) 20th European Conference on Circuit Theory and Design (ECCTD), 126-129; 
G. He, M.D. Shrimali, K. Aihara (2007) {\em International Joint Conference on 
Neural Networks}, pp. 350-354.

%	  \bibitem{May73} R. May, Ecology, {\bf 54}, 638 (1973).
 
%	 \bibitem{stability2} S. Allesina \& S.Tang, Nature, {\bf 483}, 205 (2012).
 
	%\bibitem{allee} PA Stephens, WJ Sutherland \& RP Freckleton, {\em Oikos}, 
	%{\bf 87}, 185 (1999) 
	
	%\bibitem{msd1} Amritkar, R. E., \& Rangarajan, G. (2006) {\em Phys. Rev. 
	%Letts.}, {\bf 96}, 258102.
	
	%\bibitem{msd2} Nishikawa, T., \& Motter, A. E. (2010). {\em PNAS} {\bf 
	%107}, 10342.
	
	%\bibitem{msd3} Sorrentino, F., \& Ott, E. (2008) {\em Phys. Rev. Letts.}, 
	%{\bf  100}, 114101.
 
	
	%\bibitem{balanced} N.K. Kamal and S. Sinha, {\em Chaos, Solitons and  
	%Fractals} {\bf 44} (2011) pp. 71-78
	
%	\bibitem{epjb} A. Choudhary, V. Kohar, S. Sinha, {\em Eur.  Phys. J. B} 
%{\bf 87} (2014), 202
	
	
	\bibitem{scalefree} Barabasi, A.-L. and R. Albert, {\em Science} {\bf 286}, 
	509 (1999).


	
%	\bibitem{somdatta} S. Sinha and S. Parthasarathy, {\em Proc.  Natl.  Acad.  
%Sci. }  {\bf 93}, 1504-1508 (1996).
	
	
%	\bibitem{newman} M. Newman, {\it Networks: An Introduction} (Oxford 
%University Press, Oxford, UK, 2010).
		
	\bibitem{bak} P. Bak, C. Tang and K. Wiesenfeld, {\em Phys. Rev. Letts.} 
	{\bf 59} (1987) 381.
	

	\bibitem{relax} A. Mondal and S. Sinha, {\em Phys. Rev. E} {\bf 73} (2006).
	
	\bibitem{thresh} S. Sinha, {\em Phys. Rev. E}, {\bf 63} (2001) 036212; {ibid}, 
	{\bf 69} (2004) 066209; K. Murali and S. Sinha, {\em Phys. Rev. E}, {\bf 68} 
	(2003) 016210.

 \bibitem{self-perpetuating}
 M. Scheffer, S. Rinaldi,  A. Gragnani, L. R. Mur, \& E. H. van Nes, {\em  
 Ecology}, {\bf 78}, 272 (1997).
 

\bibitem{betweeness} Betweenness centrality of a node is given as 
$b(i)=\sum_{s,t\in I}\frac{\sigma(s,t|i)}{\sigma(s,t)}$, where $I$ is the set 
of all nodes, $\sigma(s,t)$ is the number of shortest paths between nodes $s$ 
and $t$ and $\sigma(s,t|i)$ is the number of shortest paths passing through the 
node $i$.	
	
	%\bibitem{dsf} A.-L. Barabasi, E. Ravasz and T. Vicsek {\em Physica A} {\bf 
	%299} (2001) 559-564
	\end{thebibliography}
 \end{document}